\documentclass{aa}
\usepackage[varg]{txfonts}
\usepackage{hyperref}
\usepackage{subcaption} 
\usepackage{amssymb}
\usepackage[font=footnotesize]{caption}
\usepackage{nccmath}
\usepackage{placeins}

\begin{document}

\title{Assessing the detectability and identification of pulsar halos with
H.E.S.S. and CTAO}


\author{K.~Sabri\corrauth{karimsabriastro@outlook.com}
  \and Y.~Gallant\email{yves.gallant@in2p3.fr}
  \and J.~Devin\email{devin@lupm.in2p3.fr}
    }


\institute{Laboratoire Univers et Particules de Montpellier, 
Université de Montpellier, CNRS/IN2P3, Place Eugène Bataillon, 34095 Montpellier Cedex 5, France}

\date{Received *** November *** / Accepted *** January ***}

\abstract {Pulsar halos are a class of extended $\gamma$-ray sources, resulting from inverse Compton scattering of ambient photons by electrons and positrons believed to have escaped confinement in a pulsar wind nebula. First identified by HAWC towards PSR~J0633$+$1746 and PSR~B0656$+$14, a growing number of pulsar-powered $\gamma$-ray sources is proposed as candidate halos.} {Using a model of isotropic diffusion, we assessed the ability of current-and-future generation imaging atmospheric Cherenkov telescopes to characterize the physical properties of such sources in the inner Galaxy.} {We constructed template-based pulsar halo models in the Gammapy library and fitted them to simulated H.E.S.S. and CTAO data of halos with different physical parameters. The sources were characterized spectro-morphologically using 2D Gaussian components. We assessed the statistical preference between the halo and 2D Gaussian hypotheses by the simulated data and quantified the impact of the diffuse Galactic emission.} {A multitude of the halos' parameters can be constrained simultaneously, including the pulsar's true age, its proper motion, the diffusion coefficient, the index of the injected particle spectrum and the injection efficiency. In some cases, the H.E.S.S. array can statistically distinguish between the halo model and a model containing 2D Gaussian components, while a significant improvement is seen with CTAO. The source's detection significance is drastically reduced with increasing intensity of the Galactic diffuse emission. If the latter is not modeled using the data-generating model, the halo parameter estimation is biased.} {A physically-motivated model of pulsar halos can facilitate the identification and characterization of this class of sources in high-level data analyses. The simulated halos exhibit properties that may already be found in surveys of current-generation telescopes with only a modest observation time.}

\titlerunning{Detectability and identification of pulsar halos with H.E.S.S. and CTAO}
\maketitle
\nolinenumbers

\section{Introduction}
Pulsars produce emission nebulae wherein electrons and positrons ($e^{\pm}$) are confined in a nebular magnetic field originating in the pulsar. Such pulsar wind nebulae (PWNe) constitute a prominent category of very-high-energy (VHE, 100's of GeV to 10's of TeV) and ultra-high-energy (UHE, $\gtrsim10^2~\text{TeV}$) $\gamma$-ray sources (\citealt{HGPS}, \citealt{1lhaaso2024}) and are possibly important contributors to the local positron cosmic-ray (CR) spectrum. In 2017, the HAWC collaboration identified the extended $\gamma$-ray emission around the middle-aged pulsars PSR~J0633$+$1746 and PSR~B0656$+$14, hereafter the Geminga and Monogem pulsars with characteristic ages of 342~kyr and 110~kyr, respectively, as originating from inverse Compton scattering (ICS) of the ambient interstellar radiation fields (ISRFs) and the cosmic microwave background (CMB) by $e^{\pm}$ believed to have escaped the PWN \citep{HAWCGeminga17}. The age of the pulsars is such that they may have exited the supernova remnant (SNR) due to their kick velocities, and, in the case of Geminga, the extension of the VHE $\gamma$-ray emission ($\gtrsim20~\text{pc}$) is two orders of magnitude larger than that of the X-ray PWN ($\sim$ 0.1~pc) observed by the Chandra observatory (\citealt{Pavlov_2010}, \citealt{Posselt_2017}). The surface brightness profiles of both halos can be reproduced via a model of isotropic diffusion of $e^{\pm}$ in the interstellar medium (ISM), with a diffusion coefficient $D(100~\text{TeV})=4.97^{+0.97}_{-0.84}~\text{stat}~^{+4.32}_{-2.0}~\text{syst}\times10^{27}~\text{cm}^2~\text{s}^{-1}$ and $D(100~\text{TeV})=6.82^{+0.65}_{-1.11}~\text{stat}~^{+6.92}_{-3.39}~\text{syst}\times10^{27}~\text{cm}^2~\text{s}^{-1}$ for Geminga and Monogem, respectively \citep{HAWCGeminga2024}.

Interestingly, the aforementioned estimates of the diffusion coefficient are up to three orders of magnitude lower than the Galactic average if the latter is extrapolated from measurements of the local CR Boron-to-Carbon ratio \citep{ams_bc_ratio}. In the context of assessing the contribution of nearby pulsars to the local CR positron flux, a two-zone model (as opposed to the one-zone model used by \citealt{HAWCGeminga2024}) of isotropic diffusion in the ISM has been employed, featuring an inner zone extending up to 10's of pc away from the pulsar where diffusion is suppressed with respect to the assumed Galactic average, and an outer zone where it matches the average Galactic rate. Details of the model can be found in e.g. \cite{Fang_2018}, \cite{tangpiran}, \cite{DiMauro2019}, \cite{Osipov2020}, \cite{Martin2022}, \cite{Schroer2023} and references therein. Theoretically, the suppressed diffusion rate is thought to be due to self-confinement by an enhanced turbulence level induced by escaped leptons (\citealt{evoli_halo} and \citealt{mukhopadhyay_halo}) and/or by the progenitor's medium having been perturbed by the expanding SNR \citep{FangGeminga} or the collective wind activity of a massive stellar cluster. Alternatively, a model of anisotropic diffusion in the ISM is proposed by \cite{Liu2019}.

Roughly a decade of $\gamma$-ray sources studied in dedicated analyses of HAWC, LHAASO and H.E.S.S. data have been suggested to belong to this class of sources. In addition to the canonical halos of Geminga and Monogem, examples include PSR~B0540$+$23,  PSR~J0622$+$3749, PSR~J0359$+$5414, PSR J0248$+$6021 and PSR~B1055$-$52 (\citealt{HAWC_B0540}, \citealt{LHAASO_J0622}, \citealt{HAWC_J0359}, \citealt{LHAASO_J0248} and \citealt{B1055-52_proc}, respectively), with a characteristic pulsar age ranging from 62 to 538~kyr and a pulsar rotational period ranging from 79 to 385~ms. In the case of PSR~J0622$+$3749 and PSR J0248$+$6021, a spatial template of (one-zone) $e^{\pm}$ diffusion was compared with that of a 2D Gaussian through hypothesis testing, yielding inconclusive differences in the likelihoods associated to each model (based on the Akaike Information Criterion, see e.g. \citealt{aic}). Dedicated analyses of sources in the H.E.S.S. Galactic Plane Survey (HGPS, \citealt{HGPS}) catalog that are associated with powerful pulsars resolved the $\gamma$-ray emission into two components, with the larger component interpreted by means of a leptonic multi-zone transport model as originating from a population of relic $e^{\pm}$ which may have escaped the PWN, such as the case of HESS~J1809$-$193 \citep{hessj1809} and HESS~J1813$-$178 \citep{hessj1813}. In the case of HESS~J1825$-$137, one of the brightest and most extended TeV PWNe known to date and the energy-dependent morphology of which is studied in detail in \cite{hessj1825}, the spectro-morphological properties of the $\gamma$-ray emission can be reproduced with the one-zone isotropic diffusion model \citep{LiuJ1825} or with advection-diffusion models (\citealt{FangWuJ1825}, \citealt{CollinsJ1825}), while \cite{MartinEscape} showed that $e^{\pm}$ which have escaped into the ISM may represent a significant contribution to the observed $\gamma$-ray emission. In the context of detecting pulsar halos in the CTAO Galactic Plane Survey, \cite{ecknerCTAhalos} conducted a detailed study using observation simulations with the CTAO instrument response functions (IRFs) and the previously described two-zone diffusion model, finding that a substantial population of Geminga and Monogem-like halos could be detected (and possibly identified) in the future.

A template-based likelihood fit of the one-zone isotropic diffusion model, using HAWC data, was proposed in the Geminga and Monogem analysis of \cite{HAWCGeminga2024}, simultaneously yielding estimates on the diffusion coefficient, the index of the injected spectrum of $e^{\pm}$ into the ISM and the conversion efficiency of the pulsar's spin-down power into relativistic $e^{\pm}$. In the interest of facilitating the characterization and search for pulsar halos in Imaging Atmospheric Cherenkov Telescope (IACT) data, we used observation simulations with publicly available H.E.S.S. \citep{hess_public} and CTAO \citep{ctao_irfs} IRFs to demonstrate that the one-zone isotropic diffusion model can be used in the modeling framework of Gammapy \citep{gammapy:2023} to yield reliable estimates of the physical parameters. In addition to the parameters that were fitted in \cite{HAWCGeminga2024}, we show that the pulsar's true age and proper motion's transverse velocity can be simultaneously constrained. Additionally, we used the simulated H.E.S.S. and CTAO datasets to assess the ability of these instruments to characterize the spectral and morphological features of pulsar halos in dedicated analyses and within a large parameter space, and statistically distinguish the physically-motivated halo templates from simple geometric models such as 2D Gaussians, which are traditionally used to model extended pulsar-powered sources. In Section~\ref{sect:halo_mod}, the pulsar halo one-zone isotropic diffusion model, the studied parameter space and the simulation of high-level datasets with H.E.S.S. and CTAO are presented. Section~\ref{sect:simulation} describes the construction of the pulsar halo templates for fitting and the analysis method of the simulated datasets. Subsequently, illustrative cases are shown in Section~\ref{sect:illustr}. The detection and characterization of pulsar halos over the entire explored parameter space are studied in Section~\ref{sect:param_scan}, and Section~\ref{sect:GDE} discusses the impact of a Galactic diffuse emission model. We conclude with  Section~\ref{sect:conclusion}.
\section{Pulsar halo model and dataset simulation}\label{sect:halo_mod}
Electrons and positrons produced in the magnetosphere of the pulsar, which are possibly further accelerated in the wind through magnetic reconnection and/or at the wind termination shock, diffuse along magnetic field lines in the PWN, where the particle transport is primarily dominated by advection powered by the pulsar's spin-down \citep{Amato2024}. They subsequently escape into the ISM and diffuse along the ISM magnetic field lines. The halo is thought to emerge at the late stages of the pulsar-PWN-SNR system's evolution, when particle escape becomes easier either due to the disruption of the PWN by its interplay with the SNR's reverse shock, or when the pulsar has left the SNR owing to the kick velocity due to the progenitor supernova explosion's asymmetry, thereby producing a bow shock PWN \citep{giacinti_halos}. Details of the model used in this work are well documented in the literature (e.g. \citealt{Fang_2018}, \citealt{tangpiran}, \citealt{DiMauro2019}, \citealt{LiuJ1825}, \citealt{Martin2022}, \citealt{fangJ1831}, \citealt{HESSGeminga}, \citealt{Schroer2023} and references therein) and in this section we present an overview of the model with a focus on motivating the explored parameter space. Subsequently, the high-level dataset simulation method will be detailed.
\subsection{Model overview}
We adopted a model where $e^{\pm}$ are continuously injected by a PWN of negligible size (point-like) compared to the extent of the VHE emission. The $e^{\pm}$ are assumed to have escaped the PWN, diffusing in the ISM. The environment is described by an isotropic and homogeneous diffusion coefficient, a uniform magnetic field and uniform ICS target radiation fields. The transport in spherical coordinates of $e^{\pm}$ can be described by the continuity equation with radiative losses and a continuous point-like source term:
\begin{equation}
    \frac{\partial N(E,\mathbf{r},t)}{\partial t} - \mathbf{\nabla} [D(E) \mathbf{\nabla} N] - \frac{\partial }{\partial E}[\dot{E}N] = Q(E, t)\delta(\mathbf{r}-\mathbf{r_s}) \label{eq:diff_N}
\end{equation}
where $N$ is the differential particle number density at some energy $E$, time $t$ after the pulsar birth, and position $\mathbf{r}$ with $\mathbf{r_s}$ the pulsar's position. The diffusion coefficient is $D(E)=D_0\left(E/E_0\right)^{\delta}$, with $D_0$ the normalization at a reference energy $E_0$ and $\delta=1/3$ the rigidity dependence fixed according to Kolmogorov turbulence theory. Hereafter, values of $D_0$ shall be given at $E_0=100~\text{TeV}$ for direct comparison with the various aforementioned works in the literature. Synchrotron and ICS losses are assumed, with $\dot{E}=\partial{E}/\partial{t}$ the particle cooling rate. The observed synchrotron loss rate averaged over pitch angle is computed for $B_{ISM}=3~\mu\text{G}$. The loss rate from ICS is computed according to the method of \cite{Delahaye2010}, applicable to target photon fields parameterized as (diluted) black-bodies. We derived these by fitting the sum of an optical, near infrared (NIR) and far infrared (FIR) black-body, each described by a dilution factor and temperature, to the predicted spectral energy distributions of \cite{Popescu2017} at some position in the Galaxy. Throughout this work, we used the ISRFs at an arbitrary position corresponding to the Galactic coordinates $(\ell,b)=(320^{\circ}, -2^{\circ})$ and a distance from Earth of 3~kpc, and found the temperatures and energy densities shown in Table~\ref{table:isrf}.
\begin{table}[t!]
\caption{ISRF black-body components approximated from the model of \cite{Popescu2017} at $\ell=320^{\circ}$, $b=-2^{\circ}$ and a distance from Earth of 3~kpc.}              
\label{table:isrf}      
\centering                                      
\begin{tabular}{c c c}          
\hline\hline                        
Component & Temperature & Energy density\\    
\hline                                   
    Optical & 3310~K & 1.2~eV~cm$^{-3}$ \\      
    NIR & 550~K & 0.3~eV~cm$^{-3}$\\
    FIR & 40~K & 0.5~eV~cm$^{-3}$ \\
\hline                                             
\end{tabular}
\end{table}
\begin{table*}[t!]
\caption{Parameter values of the simulated systems, compared to those of the Geminga and Monogem halos from \cite{HAWCGeminga2024}. The upper section corresponds to parameters for which multiple values were simulated.}              
\label{table:sim}      
\centering                                      
\begin{tabular}{c c c c}          
\hline\hline                        
Parameter &Geminga &Monogem& This work \\    
\hline                                   
    $P_0$~[ms]&-&-&30; 70\\
        $V_{PSR}$~[km~s$^{-1}$] &211\tablefootmark{a}&59\tablefootmark{b}& 0; 300; 950\\
        $D_0$~[cm$^{2}$~s$^{-1}$]&$4.97^{+0.97}_{-0.84}~\text{stat}~^{+4.32}_{-2.0}~\text{syst}\times10^{27}$&$6.82^{+0.65}_{-1.11}~\text{stat}~^{+6.92}_{-3.39}~\text{syst}\times10^{27}$& $10^{26}$ to $10^{28}$ (6 points)\\
         $\Gamma$&$0.95^{+0.08}_{-0.09}~\text{stat}\pm0.32~\text{syst}$&$1.06^{+0.09}_{-0.10}~\text{stat}\pm0.14~\text{syst}$& 1; 2\\
         $E_c$~[TeV]&$10^2$&$10^2$&$10^2$; $10^3$\\
         $\eta$&$6.60\pm0.7~\text{stat}\pm2.5~\text{syst}~\%$&$5.10^{+0.70}_{-0.80}~\text{stat}\pm2.7~\text{syst}~\%$&$3\%$\\
         \hline
         $V_{PSR, \Theta}$~[$^{\circ}$] &-&-& 180\\
         $d_{PSR}$~[kpc]&0.25&0.29&3\\ 
         $\tau_c$~[kyr]&342&110&100\\
         $n$&$3$&$3$&3\\
         $P$~[ms]&237&385&80\\
         $B_{ISM}$~[$\mu$G]&$3$&$3$&$3$\\
         $\delta$&1/3&$1/3$&$1/3$\\
\hline                                              
\label{tab:model_vals}
\end{tabular}
\tablefoot{
\tablefoottext{a}{\cite{Faherty2007}}
\tablefoottext{b}{\cite{hobbs_ppm}}}
\end{table*}

The pulsar is treated as a rotating electromagnetic dipole following e.g. \cite{lorimer_book}. The true age of the pulsar $t_{age}$ is a function of the present-day period $P$, its evolution $\dot{P}$, the initial period $P_0$, and the braking index $n\neq1$ as follows:
\begin{equation}
    t_{age}=\frac{P}{(n-1)\dot{P}}\left(1-\left(\frac{P_0}{P}\right)^{n-1}\right)
\end{equation}
Assuming a pure electromagnetic dipole, $n=3$ such that the characteristic age is $\tau_c=\tau_0+t_{age}$ with $\tau_0$ the initial spin-down timescale. The pulsar's spin-down evolution with some time $t$ after the pulsar's birth reads: 
\begin{equation}
    L(t)=L_{\star}\left(\frac{1+t_{age}/\tau_0}{1+t/\tau_0}\right)^2
\end{equation}
with $L_{\star}=L(t_{age})\propto P^{-2}\tau_c^{-1}$ its present-day spin-down power. The field of view of H.E.S.S. is on the order of the full extent of the Geminga and Monogem halos detected by HAWC. It follows that similar halos are more likely to be detected in H.E.S.S. data around pulsars farther from Earth than Geminga and Monogem ($d_{PSR}=250$~pc and 288~pc, respectively). This, however, would require that they have a higher spin-down luminosity than those of Geminga and Monogem ($3.2\times10^{34}~\text{erg}~\text{s}^{-1}$ and $3.8\times10^{34}~\text{erg}~\text{s}^{-1}$, respectively). It was shown by \cite{carrigan} and \cite{HESS_PWNe} that the number of spatial coincidences between powerful pulsars with $L_{\star}/d_{PSR}^2\geq10^{34}~\text{erg}~\text{s}^{-1}~\text{kpc}^{-2}$ and VHE components in H.E.S.S. data is significantly higher than the expected number of chance coincidences, meaning that such pulsars are generally able to produce detectable VHE emission. In this work, we considered a hypothetical pulsar at $d_{PSR}=3~$kpc, meaning that $L_{\star}$ must be roughly an order of magnitude higher than that of Geminga and Monogem. We assumed $\tau_c=100~\text{kyr}$ and $P=80~\text{ms}$, resulting in $L_{\star}=9.8\times10^{35}~\text{erg}~\text{s}^{-1}$.

The particle injection term $Q(E,t)$, which follows a power law (PL) with exponential cutoff $E_c$ and index $\Gamma$, is normalized such that the total energy injected in $e^{\pm}$ per unit time is some fraction $\eta$ (the injection efficiency) of the pulsar's spin-down power:
\begin{equation}
    Q(E,t)=Q_0(t)\left(\frac{E}{E_0}\right)^{-\Gamma}\exp\left(\frac{-E}{E_c}\right)
\end{equation}
\begin{equation}
    Q_0(t)=\frac{\eta L(t)}{\int_{E_{min}}^{E_{max}} E\left(\frac{E}{E_0}\right)^{-\Gamma}\exp\left(\frac{-E}{E_c}\right)\,dE} \label{eq:q_norm}
\end{equation}
with a reference energy $E_0=1~$TeV, minimum energy $E_{min}=0.1$~TeV and maximum energy $E_{max}=1$~PeV.

The ICS $\gamma$-ray intensity $I(E_{\gamma}, \ell, b)$ is computed using the Naima package \citep{naima} after projecting the differential particle density distribution $N$ onto the sky. The semi-analytical solution to Eq.~\ref{eq:diff_N} including the pulsar's proper motion $\mathbf{V}_{PSR}$ and the projection are detailed in Appendix~\ref{app:halo_model}. The pulsar's proper motion direction in the sky is parametrized through a rotation of the intensity about the pulsar's present-day position by an angle $V_{PSR, \Theta}$, taken counter clock-wise from the increasing Galactic longitude direction. Hereafter, $V_{PSR, \Theta}=180^{\circ}$ such that the pulsar is traveling horizontally towards decreasing longitudes. In \cite{Zhang_2021}, it was found that the proper motion's line-of-sight component had a negligible effect on the halo's morphology if $d_{PSR}$ is much larger than the distance traveled by the pulsar since birth. The distances traveled by middle-aged pulsars are on the orders of tens to hundreds of parsecs at most, and the line-of-sight component is therefore neglected in this work.
\subsection{Simulated parameter space}
We simulated observations of halos with different model parameter values, as shown in Table~\ref{table:sim}, for a total of 144 different parameter value combinations. Additionally, Table~\ref{table:sim} contains the fitted and fixed values reported in ~\cite{HAWCGeminga2024} for the Geminga and Monogem halos, along with their pulsars' properties.

The age of the system is expected to influence the spectral and morphological properties of the $\gamma$-ray emission observed with IACTs. Therefore, we considered two values of the hypothetical pulsar's initial period $P_0$, with $P_0=30~$ms corresponding to an older case with true age $t_{age}\sim85~$kyr, and $P_0=70~$ms corresponding to a younger case with $t_{age}\sim25~$kyr. The latter is significantly younger than identified and candidate pulsar halos and it is likely that in a more realistic model, the medium in which $e^{\pm}$ are diffusing is not characteristic of the ISM (e.g. \citealt{Bourguinat_2026}). Moreover, the $\gamma$-ray emission features are more sensitive to the early injection history of $e^{\pm}$ and the onset of the halo formation, which are neglected in the model used in this work. Nevertheless, \cite{LiuJ1825} have shown that the model can reproduce the features observed in HESS~J1825$-$137 with $t_{age}\sim30~$kyr, which motivates the inclusion of the $t_{age}\sim25~$kyr scenario in our explored parameter space. We also considered cases where the pulsar has a non-zero proper motion with $V_{PSR}=300~\text{km}~\text{s}^{-1}$ and $V_{PSR}=950~\text{km}~\text{s}^{-1}$, because it could have an influence on the $\gamma$-ray morphology \citep{Zhang_2021}. The distribution of known pulsars' transverse velocity peaks between 300-400~km~s$^{-1}$ and $V_{PSR}\sim10^3~$km~s$^{-1}$ is a rare occurrence in reality (e.g. \citealt{hobbs_ppm}), but this case was included in our work for completeness.

For the diffusion coefficient normalization, 6 different values were simulated, ranging from $D_0=10^{26}~\text{cm}^2~\text{s}^{-1}$ to $10^{28}~\text{cm}^2~\text{s}^{-1}$. The highest three values scanned in this work are representative of the Geminga and Monogem halos within uncertainties. Higher values of the diffusion coefficient have not been explored because the source becomes undetectable in the H.E.S.S. observation simulations, as will be shown in Section~\ref{sect:param_scan}. Presently, the lowest value measured for a pulsar halo candidate was found by \cite{B1055-52_proc} for PSR~B1055$-$52 with $D_0=\left(3.2\pm1.2\right)\times10^{26}~\text{cm}^2~\text{s}^{-1}$ and it depends on the assumed magnetic field intensity and distance to the pulsar. For the magnetic field intensity of $3~\mu\text{G}$ assumed in this work, the PSR~B1055$-$52 diffusion coefficient is well below the Bohm limit derived from quasi-linear theory and would require levels of turbulence $\delta B_{ISM}/B_{ISM}\gg1$ (see e.g. \citealt{Hussein_2014} and references therein). The likelihood of encountering the scenario in nature is outside the scope of this work and the inclusivity of the scanned parameter space was prioritized.

Two values of the spectral injection index were tested, with $\Gamma=1$ and $\Gamma=2$. The former value is consistent with the HAWC measurements of the Geminga and Monogem halos (cf.~Table~\ref{tab:model_vals}). We included $\Gamma=2$ in the parameter space, which is also used in \citealt{Zhang_2021}, because the injection spectra corresponding to young PWNe exhibit indices between 2.2 and 2.8 above a few tens of GeV (e.g. \citealt{bucciantini_pwne} and \citealt{torres_pwne}). Additionally, we tested two values of the injection cut-off energy with $E_c=100~$TeV and $E_c=1~$PeV, both of which are within the limits of the maximum achievable energy of relativistic $e^{\pm}$ accelerated by young pulsars \citep{dOW_uhe_pulsars}. A harder injection spectrum implies a higher proportion of $e^{\pm}$ injected at higher energies, and thus the $\gamma$-ray SED becomes more sensitive to variations in the injection cutoff energy. For a higher $E_c$, more of the pulsar's rotational power goes into the injection of particles that will contribute less $\gamma$ rays in the H.E.S.S. energy band. This results in a less significant source for which it is difficult to study the observational impact of varying $E_c$. As such, we fixed the injection efficiency $\eta$, which acts as a normalization, to $3\%$ in all cases except for $\Gamma=1$ and $E_c=1~$PeV, where it was increased to $7\%$ such as the significance of the source remains roughly similar to the $E_c=100~$TeV case. The chosen values of $\eta$ are similar to the measurements of \cite{HAWCGeminga2024} for the Geminga and Monogem halos (cf.~Table~\ref{tab:model_vals}). We note that, unless otherwise stated, the analysis results of the subsequently simulated datasets are highly similar between the $E_c=100~$TeV and $1~$PeV cases, and thus only the former shall be explored in this paper.

\begin{figure*}[t!]
\centering
   \includegraphics[width=17cm]{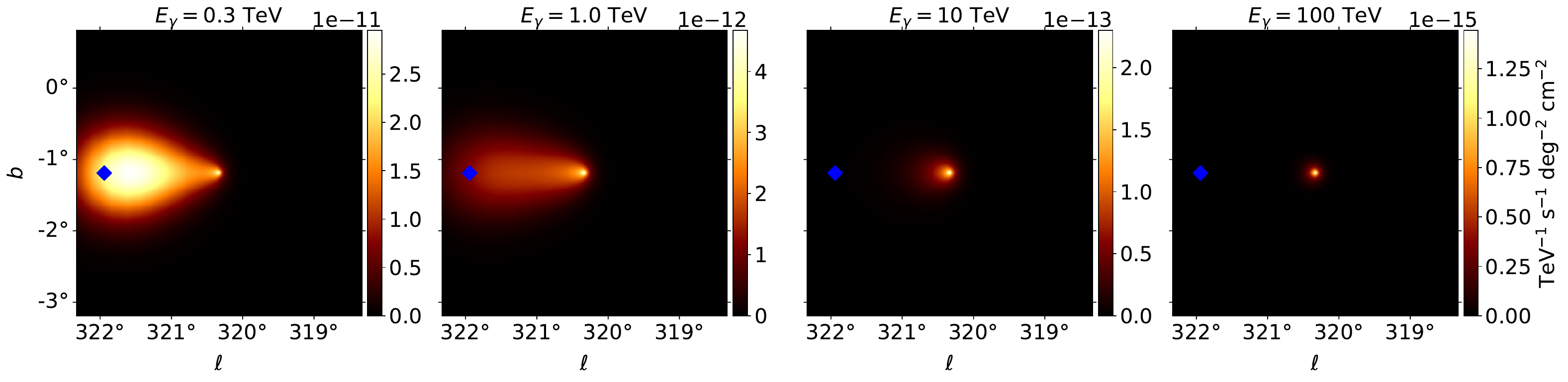}
     \caption{Intensity maps $I(E_{\gamma}, \ell, b)$ of an example pulsar halo model used for the simulations. The model parameters are $D_0=2\times10^{27}~$cm$^2$~s$^{-1}$, $V_{PSR}=950~$km~s$^{-1}$, $\Gamma=1$, $P_0=30~$ms and $\eta=3\%$. The pulsar is traveling towards decreasing longitudes, with $V_{PSR, \Theta}=180^{\circ}$. The values of other model parameters are shown in Table~\ref{tab:model_vals}. The pulsar's birth position is shown as a blue diamond, and its present-day position is the center of the field of view.}
     \label{fig:example_intensity_map_ppm}
\end{figure*}

To illustrate the energy-dependent morphology of the halo, maps of the model intensity $I(E_{\gamma}, \ell, b)$ are shown at different photon energies in Fig.~\ref{fig:example_intensity_map_ppm} for $V_{PSR}=950~\text{km}~\text{s}^{-1}$. A prominent "tail-like" structure of $\gamma$ rays from relic particles is predicted at lower energies, a feature that was studied in detail by \cite{Zhang_2021}. In the left-most panel at $E_{\gamma}=0.3~\text{TeV}$, a double-peaked structure is seen, with $\gamma$ rays concentrated near the present-day pulsar position in addition to a secondary, more extended structure near the pulsar's birth position. At $E_{\gamma}=1~\text{TeV}$, the secondary structure near the pulsar's birth location is less pronounced, and fades at higher energies where $e^{\pm}$ are concentrated near the pulsar's position.

\begin{figure*}[t!]
 \centering
    \includegraphics[width=17cm]{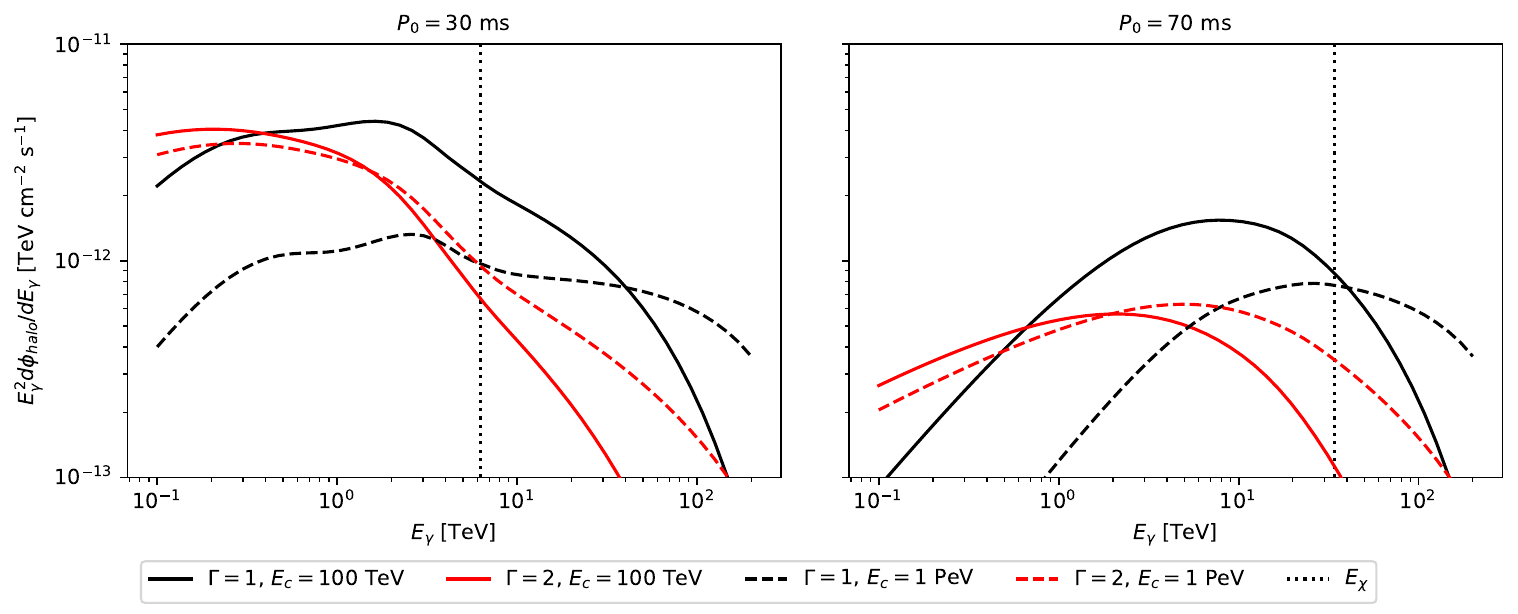} 
  \caption{\textbf{Left:} $\gamma$-ray SEDs of the simulated pulsar halo model with $P_0=30~$ms ($t_{age}=85~$kyr). \textbf{Right:} $\gamma$-ray SEDs of the simulated pulsar halo model with $P_0=70~$ms ($t_{age}=25~$kyr). The black lines represent a harder injection spectrum with $\Gamma=1$, and the red lines represent a softer $\Gamma=2$. The solid lines correspond to an injection cutoff $E_c=100~$TeV and the dashed lines to $E_c=1~$PeV. The vertical dotted lines indicate $E_{\chi}$, the $e^{\pm}$ energy for which the cooling timescale $t_{cool,max}$ is equal to $t_{age}$. The $e^{\pm}$ injection efficiency is $\eta=3\%$.}
  \label{fig:model-spectra}
\end{figure*}

Fig.~\ref{fig:model-spectra} shows the simulated models' $\gamma$-ray spectral energy distributions (SEDs) for different values of $P_0$, $\Gamma$ and $E_c$. With an assumed uniform magnetic field, the diffusion coefficient normalization $D_0$ and the pulsar's proper motion $\mathbf{V}_{PSR}$ have a negligible impact on the SED since the differential photon spectrum $d\phi_{halo}/dE_{\gamma}$ is computed by integrating the intensity $I(E_{\gamma}, \ell, b)$ over the entire field of view, a region extending farther than the diffusion length scale and the distance traveled by the pulsar since birth. The photon SED in the older pulsar case is shown in the left panel of Fig.~\ref{fig:model-spectra}, and that of the younger case is shown in the right panel. The $e^{\pm}$ energy for which the cooling timescale $t_{cool,max}$ is equal to the pulsar's true age, $E_{\chi}$ as defined in Appendix~\ref{app:halo_model}, is shown as a dotted line in each panel. Observed $\gamma$ rays below this energy are mainly produced by a ``relic'' population of $e^{\pm}$ which were injected throughout the pulsar's history. Observed $\gamma$ rays above this energy can only be produced by the most recently injected $e^{\pm}$. Substantially less relic $e^{\pm}$ are injected in the younger pulsar case than in the older case, resulting in a significantly weaker photon spectrum below $E_{\chi}$. The halo is also expected to be more extended in the older case (for the same velocity $V_{PSR}$), since relic $e^{\pm}$ diffuse farther away than the more recently injected population.
\subsection{High-level data simulation}\label{subsect:halo_simulation}
With real data, analyses of pulsar halos near the Galactic plane are possibly subject to source confusion. Thus, it is appropriate to adopt the spectro-morphological (i.e. 3D) analysis method, wherein the spectral and spatial components of an intensity model are constrained simultaneously through a likelihood fit to a 3D binned events count cube. We used version 1.2 of the Gammapy library \citep{gammapy:zenodo-1.2} to simulate high-level datasets of H.E.S.S. and CTAO observations of the previously described pulsar halo model.

The first H.E.S.S. public data release\footnote{https://doi.org/10.5281/zenodo.1421099} \citep{hess_public} includes $\sim8.4$ hours of observations targeting the composite SNR MSH~15$-$52 using the array's CT1 to CT4 telescopes. Additionally, to each observation in the H.E.S.S. data release corresponds a 3D model of the residual hadronic CR shower background rate, built by \cite{FoVbkgMohrmann} from observations of extra-galactic regions devoid of significant $\gamma$-ray emission. We used the IRFs of the MSH~15$-$52 observations because they are representative of an observation strategy dedicated to an individual extended source near the Galactic plane (in terms of wobble offset and the position of the source in the sky). We replicated the MSH~15$-$52 observations 5 times to get a total of $\sim42$ hours of observation. To simulate the reconstructed events from one observation, we used its pointing information, duration and IRFs (point-spread function, energy dispersion, and effective detector area) with the pulsar halo model, assuming the pulsar's present-day position is that of MSH~15$-$52. Additionally, events from the observation's corresponding background model were generated. The events were binned in a $4^{\circ}\times4^{\circ}$ counts cube centered on the position of MSH~15$-$52 with a spatial bin size of $0.02^{\circ}$ and a reconstructed energy axis from 0.3 to 100 TeV (8 bins per decade). The true energy axis used by the IRFs is defined from 0.1 to 200 TeV (20 bins per decade). The binned event counts were then Poisson-fluctuated. The reconstructed events were selected such that they lie within $2^{\circ}$ of the camera center, with an energy such that the effective detector area is at least $10\%$ of the maximum effective area, and which is at least higher than the hadronic CR background model peak energy. Through a maximum likelihood fit, the background model is re-adjusted with a PL in energy (with free normalization) to the binned events outside a $4^{\circ}\times2^{\circ}$ rectangular mask along the middle of the field of view. This was done for the entire set of observations, and all the generated datasets were stacked together. We note, as stated in \citep{hess_public}, that the IRFs used in this work do not reflect on the H.E.S.S. performance with state-of-the-art event reconstruction and selection methods (see e.g. \citealt{deNaurois_model}, \citealt{parsons_impact},  \citealt{Khelifi_hapfr} and references therein).

The configuration for the simulation of datasets using the CTAO IRFs\footnote{https://doi.org/10.5281/zenodo.5499840} is the same as in the H.E.S.S. case, except that events were selected within $3.5^{\circ}$ of the camera center instead of $2.0^{\circ}$. The IRFs corresponding to the Alpha layout of Prod 5 v0.1 \citep{ctao_irfs} were used. Specifically, we used the CTAO-South's Medium-Sized Telescope (MST) sub-array (which consists of 14 MSTs) optimized for 50 hours, and a zenith angle of $40^{\circ}$. The azimuth-averaged IRFs were used. This release includes simulated residual hadronic background events which allows for creating and fitting background models as described previously for the H.E.S.S. analysis. We note that we found similar results when using the full Alpha configuration, i.e. 14 MSTs and 37 Small-Sized Telescopes, with the exception that the simulated source's statistical detection significance was up to $\sim50\%$ higher. Thus, we only show results using the MST sub-array in this work.

For each of the 144 parameter combinations (cf. Table~\ref{tab:model_vals}), we simulated 160 datasets with both the H.E.S.S. and CTAO IRFs as described above, each referred to as a "realization".
\section{Halo template fitting and data analysis method}\label{sect:simulation}
\subsection{Construction of the halo templates for model fitting and method validation}\label{subsect:halo_template}
The pulsar halo model can be fitted directly to high-level data in a spectro-morphological data analysis, by creating spatial and spectral templates of the $\gamma$-ray emission and using them with the Gammapy modeling framework. To that end, assuming the fixed parameters in Table~\ref{tab:model_vals}, the model $\gamma$-ray intensity $I(E_{\gamma}, \ell, b)$ was computed for different combinations of the values of the parameters to be fitted ($D_0$, $\Gamma$, $\eta$, $V_{PSR}$, $V_{PSR,\Theta}$ and $P_0$). The intensity is proportional to $\eta$ and this term acts as a normalization of the model during the fit. The spatial templates for different pulsar proper motion directions can be obtained by rotating the model intensity by $V_{PSR,\Theta}$ about the pulsar's present-day position, and thus it is unnecessary to compute the intensity for different values of $V_{PSR,\Theta}$. The model intensity was computed over the following free parameter grids:
\begin{itemize}
  \item $D_0$: geometric grid from $D_{0,\text{min}}=3\times10^{25}$~cm$^2$s$^{-1}$ to $D_{0,\text{max}}=1\times10^{29}$~cm$^2$s$^{-1}$ with 21 points;
  \item $\Gamma$: linear grid from $\Gamma_{\text{min}}=0.5$ to $\Gamma_{\text{max}}=3.0$  with 17 points;
  \item $V_{PSR}$: linear grid from $V_{PSR,\text{min}}=10~\text{km}~\text{s}^{-1}$ to $V_{PSR,\text{max}}=2000~\text{km}~\text{s}^{-1}$ with 17 points;
  \item $P_{0}$: linear grid from $P_{0, \text{min}}=10$~ms to $P_{0, \text{max}}=P$ with 18 points (the model is null at $P_{0, \text{max}}$). 
\end{itemize}
The number of points for the grids was chosen so the subsequent linear interpolation of the templates, and the fit of the model to a simulated pulsar halo dataset, yielded accurate results (see the validation method at the end of this subsection and Appendix~\ref{app:param-rec}) with a reasonable computation time. Importantly, the parameter values of the simulated halos (cf. Table~\ref{tab:model_vals}) were chosen such that they are not among the aforementioned points, allowing us to test for uncertainties on the halo parameter estimation introduced by the linear interpolation. We note that the pulsar's proper motion ($V_{PSR}$, $V_{PSR,\Theta}$) and the injection index $\Gamma$ are not fitted simultaneously for computational reasons. In the likelihood fitting process, an expansive parameter space is scanned, where at each step a linear interpolation of the halo spectral and spatial templates is computed. If the number of parameters is too high, the fitting process can take an excessive amount of time (and computational resources). Moreover, the intensity model with non-zero pulsar proper motion contains two spatial dimensions instead of one in the null proper motion case (cf.~Appendix~\ref{app:halo_model}), and is slower to compute over all value combinations of the free parameter grids. Thus, when the proper motion is used, the model is only computed for one value of $\Gamma$ and the latter is not fitted. We also note that the pulsar's position is always fixed to the simulated pulsar's position (i.e. that of MSH~15$-$52) and does not count as a free parameter, since it is assumed that the halo is associated to a known pulsar. 

The Gammapy modeling framework for spectro-morphological fitting employs intensity models factorized as a spatial component (in units of steradians$^{-1}$) and a spectral component (in units of TeV$^{-1}$~cm$^{-2}$~s$^{-1}$). Consequently, a template of the differential photon spectrum $d\phi_{halo}/dE_{\gamma}=\iint d\ell db I$ (the spectral component) and the corresponding normalized spatial template $I_{\text{norm}}$ (the spatial component) are computed for all combinations of the free parameter grid values. The spatial template is obtained by normalizing the intensity in the filed of view, i.e. $I_{\text{norm}}=I/(d\phi_{halo}/dE_{\gamma})$. Both components are evaluated by linear interpolation in the free parameter grids. The Gammapy fitting backend handles the computation of the model-predicted event counts in each bin of the dataset cube. 

For each of the simulated datasets (i.e. realizations) described in Section~\ref{subsect:halo_simulation}, the pulsar halo template model was fitted to the simulated dataset using a maximum likelihood fit, and the estimates of the free parameters were recorded. Model parameters that are not free are fixed to those of the simulated halo. Additionally, during the fits the background model is re-adjusted with a PL in energy with free normalization. A realization is rejected if the fit of the pulsar halo template did not converge. For each simulated parameter combination, we checked that the simulated halo's parameters lie within one standard deviation of the average estimated values across all corresponding realizations. An example with the H.E.S.S. IRFs is provided in Appendix~\ref{app:param-rec}. The halo template fit yields reliable parameter estimates, validating the method.
\subsection{Alternative models}\label{subsect:gaussian_modeling}
In this work, we assess the detection significance of the hypothetical pulsar halos, and the ability of H.E.S.S. and CTAO to distinguish them from simpler spatial geometric models such as 2D Gaussians, which have been traditionally used to model extended pulsar-powered sources. To that end, we conducted hypothesis tests, for all realizations, between the fitted pulsar halo template models and models of 2D spatial Gaussian components with a PL or log-parabola (LP) spectrum. The spatial 2D Gaussian component is expressed as:
\begin{equation}
    G(\ell, b)=\frac{1}{2\pi\sigma^2}\exp\left(-\frac{1}{2}\frac{\theta^2}{\sigma^2}\right)
\end{equation}
with $\theta$ the angular separation between the Gaussian's centroid $(\ell_0, b_0)$ and $(\ell, b)$. If the 2D Gaussian is "elliptical", then $\sigma$ is replaced with
\begin{equation}
    \sigma_{eff}(\ell, b)=\sqrt{(\sigma_M\sin{(\Delta\phi)})^2+(\sigma_m\cos{(\Delta\phi)})^2}
\end{equation}
with $\sigma_M$ and $\sigma_m$ the major and minor semi-axes respectively, and $\Delta\phi$ the difference in position angles between $(\ell_0, b_0)$ and $(\ell, b)$. The spectral component can be modeled as a PL:
\begin{equation}
    \frac{d\phi}{dE_{\gamma}}=\phi_0\left(\frac{E_{\gamma}}{E_{0,\gamma}}\right)^{-\alpha}
\end{equation}
or with a curvature as a spectral LP:
\begin{equation}
    \frac{d\phi}{dE_{\gamma}}=\phi_0\left(\frac{E_{\gamma}}{E_{0,\gamma}}\right)^{-\alpha-\beta \log\left(E_{\gamma}/E_{0,\gamma}\right)}
\end{equation}
The reference energy is $E_{0,\gamma}=1~\text{TeV}$ and $\phi_0$ is the amplitude with units $\text{TeV}^{-1}~\text{cm}^{-2}~\text{s}^{-1}$.

Analyses of extended sources near the Galactic plane can be strongly impacted by modeling of the large-scale Galactic diffuse emission (GDE) due to (mostly) pion decay. Simulations of datasets that include such a component were done to assess the impact of the GDE on the results, and these will be discussed in Section~\ref{sect:GDE}.
\subsection{Analysis procedure}\label{subsect:analysis_method}
Throughout this work, the test statistic $\text{TS}$ associated to some model $\mathcal{M}$ is defined as $\text{TS}=-2\ln{\mathcal{L}(\mathcal{M})}$ (i.e. the Cash statistic, \citealt{cash}) with $\mathcal{L}(\mathcal{M})$ the maximum likelihood of model $\mathcal{M}$. For each realization, the pulsar halo template model was fitted onto the simulated dataset (cf.~Section~\ref{subsect:halo_template}) to compute its associated test statistic (hereafter $\text{TS}_{\text{halo}}$). The fitted parameters are $D_0$, $P_0$, $\eta$ and either $\Gamma$ or $V_{PSR}$ and $V_{PSR,\Theta}$, i.e. the model has 4 or 5 degrees of freedom (d.o.f.) depending on whether $\Gamma$ or the proper motion is free (i.e. whether or not the simulated pulsar has a null proper motion). Subsequently, the halo template model was replaced with an alternative model consisting of a fitted 2D Gaussian spatial component (3 d.o.f.) and PL spectrum (2 d.o.f.), hereafter the "Gaussian model", and its associated test statistic was computed. The detection threshold is a difference of 37 in the test statistic vs. the background-only hypothesis, which corresponds to a $5\sigma$ detection for 5 d.o.f. If the source was significantly detected, a second 2D Gaussian~$\times$~PL component is added to the model and fitted (adding 5 d.o.f. to the Gaussian model). The second Gaussian was accepted if it further improved the test statistic by a difference of 37. The eccentricity and position angle of the larger 2D Gaussian were then freed if the simulated pulsar halo had a non-zero proper motion, adding 2 d.o.f. to the Gaussian model. The ellipticity of the second Gaussian was accepted if it improved the test statistic by a difference of 12, corresponding to the $3\sigma$ threshold for 2 d.o.f. If the symmetric second Gaussian was not detected initially, the elliptical second Gaussian was accepted if it improved the test statistic by a difference of 42 vs. the single Gaussian case (corresponding to the $5\sigma$ threshold for 7 d.o.f.). A curvature in the spectra of the Gaussian model's components was then tested, by replacing the PL spectrum with a LP spectrum. If the model had one component, then the LP hypothesis was accepted if it improved the test statistic by a difference of 9 ($3\sigma$ for 1 d.o.f.). If the model had two components, then three hypotheses were tested: both components have a curved spectrum ($\text{LP}_{\text{both}}$), the first one has a curved spectrum ($\text{LP}_{\text{1}}$), or the second one has a curved spectrum ($\text{LP}_{\text{2}}$). $\text{LP}_{\text{both}}$ was accepted if $\text{TS}_{\text{LP}_{\text{both}}}-\text{TS}_{\text{LP}_{\text{1}}}\geq9$ and $\text{TS}_{\text{LP}_{\text{both}}}-\text{TS}_{\text{LP}_{\text{2}}}\geq9$. If $\text{LP}_{\text{both}}$ was not accepted, but $\text{LP}_{\text{1}}$ and $\text{LP}_{\text{2}}$ were both significant vs. their PL counterparts, then only the hypothesis with the best test statistic was accepted. All of the components' parameters were kept free and fitted simultaneously. During all fits, the background model was re-adjusted using a PL in energy with free normalization.

The Gaussian and pulsar halo models are not nested and do not satisfy the conditions of Wilk's theorem. To determine whether the Gaussian model can be rejected in favor of the halo template model, we used the Akaike Information Criterion (AIC) by computing $\Delta \text{AIC}=\text{AIC}_{\text{Gauss}}-\text{AIC}_{\text{halo}}$, with $\text{AIC}_{\text{halo}}$ and $\text{AIC}_{\text{Gauss}}$ the AIC scores of the fitted halo and Gaussian models respectively. A negative $\Delta \text{AIC}$ implies that the 2D Gaussian model has a better (lower) AIC score than that of the halo model.

\section{Morphological and spectral features with H.E.S.S. and CTAO}\label{sect:illustr}
The analysis results of individual, arbitrarily chosen realizations are illustrated in this section to highlight the detected morphological and spectral features of the simulated halo's emission for different parameter values. The analysis procedure described for one realization (cf. Section~\ref{subsect:analysis_method}) was done with both the H.E.S.S. and CTAO IRFs for all realizations and all simulated parameter combinations (cf. Table~\ref{tab:model_vals}) and the results will be shown in Section~\ref{sect:param_scan}.
\subsection{H.E.S.S. sample realizations}
\begin{figure*}
 \centering
    \includegraphics[width=15.5cm]{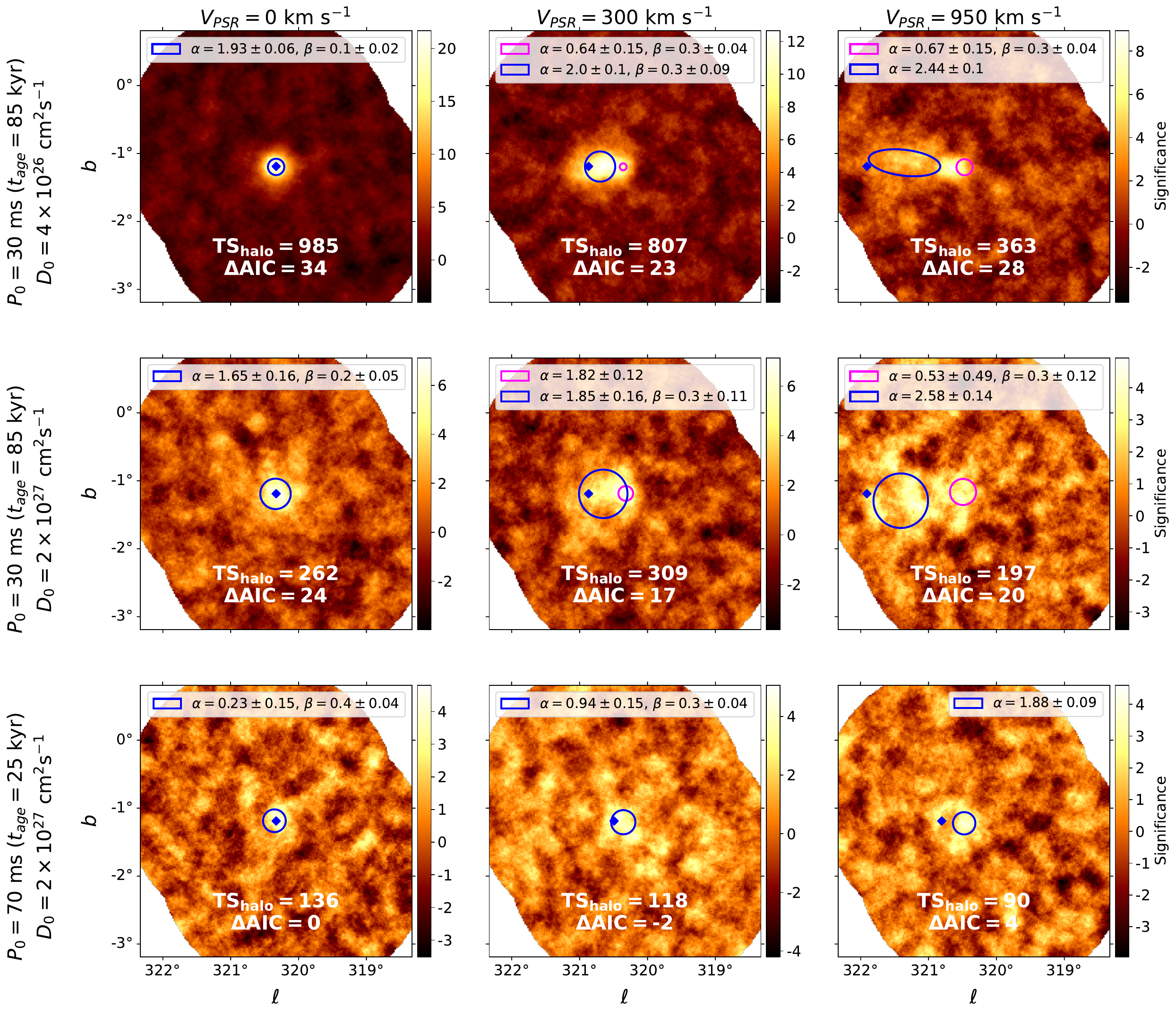} 
  \caption{Example significance maps of simulated datasets (each panel corresponds to a realization) with $\sim42$ hours of observation overlaid with the best-fit Gaussian components' $39\%$ containment region, for different values of the proper motion $V_{PSR}$, diffusion coefficient normalization $D_0$ and initial period $P_0$. In all the panels, $\Gamma=1$. The pulsar's birth position is shown as a blue diamond. $\text{TS}_{\text{halo}}$ corresponds to the test statistic of the fitted halo model. $\Delta \text{AIC}$ indicates the difference in the $\text{AIC}$ score between the overlaid model and the pulsar halo hypothesis. If two Gaussian components are found to be significant, the more compact component is shown in magenta.}
  \label{fig:hess-maps-all}
\end{figure*}

Fig.~\ref{fig:hess-maps-all} shows the significance maps of some realizations corresponding to different values of the diffusion coefficient normalization $D_0$, the pulsar's proper motion velocity $V_{PSR}$ and its initial period $P_0$, with a spectral injection index $\Gamma=1$. The maps represent the significance (in Gaussian $\sigma$) of the $\gamma$-ray excess above the background level within a correlation radius of $0.15^{\circ}$ in the 0.3-100~TeV energy band. The best-fit model containing 2D Gaussian components is overlaid as the spatial components' $39\%$ containment region, and the indices of the spectra are indicated in each panel's legend. If the LP hypothesis was significantly preferred over the PL, then the $\beta$ parameter is indicated. The test statistic of the best-fit halo model, $\text{TS}_{\text{halo}}$, is indicated along with the $\Delta \text{AIC}$ between it and the overlaid model.
\begin{figure*}
 \centering
    \includegraphics[width=15.5cm]{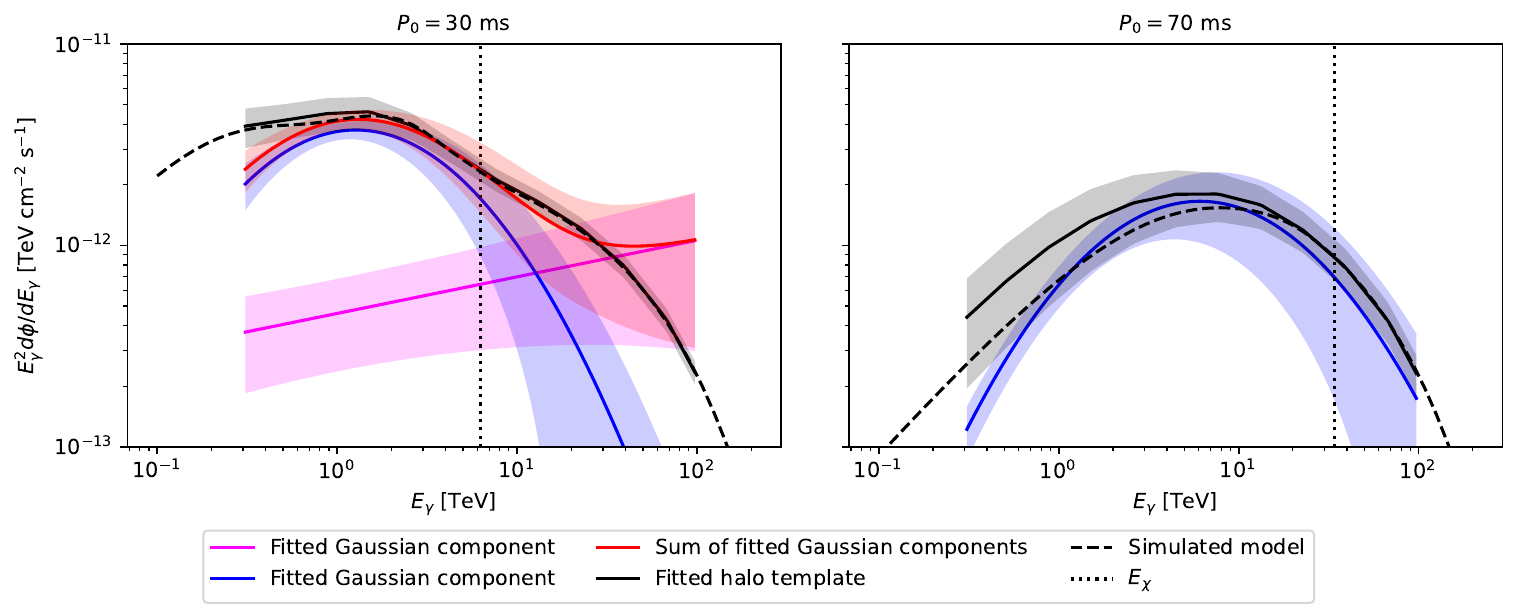} 
  \caption{SED of the simulated pulsar halo (dashed black lines) together with that of the fitted components (solid lines) obtained with 42 hours of H.E.S.S. observations. The simulated parameters are $D_0=2\times10^{27}~$cm$^2$~s$^{-1}$, $V_{PSR}=300~$km~s$^{-1}$and $\Gamma=1$. The left panel shows the older pulsar halo's SED ($t_{age}=85~\text{kyr}$) and the right panel that of the younger one ($t_{age}=25~\text{kyr}$). The dashed lines correspond to the spectra of the simulated pulsar halos.  If the Gaussian model includes two components, their sum is shown as a red line. The error bands represent the $68\%$ confidence interval.}
  \label{fig:hess-spectra_high_d}
\end{figure*}

The first row shows realizations for an initial pulsar period $P_0=30~$ms (corresponding to a true age of 85~kyr) and a diffusion coefficient normalization $D_0=4\times10^{26}~\text{cm}^2~\text{s}^{-1}$. A second Gaussian component was significantly detected in the cases with $V_{PSR}>0$~km~s$^{-1}$. The more compact component in both of these realizations exhibits a significantly harder $\gamma$-ray spectral index ($\alpha\sim0.6$) than the larger component ($\alpha\sim2.2$). The former accounts for recently injected $e^{\pm}$, while the latter accounts for the more diffuse contribution by relic $e^{\pm}$ stretching towards the pulsar's birth position. At the highest velocity, the ellipticity of the relic component significantly improved the fit and was required to reproduce the high spatial asymmetry of the emission. A curvature in the spectrum was significantly detected for all cases except the relic component of the $V_{PSR}=950~$km~s$^{-1}$ case. The test statistic of the halo model hypothesis, $\text{TS}_{\text{halo}}$, globally decreases with increasing $V_{PSR}$ due to the lower surface brightness, from $\text{TS}_{\text{halo}}=985$ at $V_{PSR}=0~$km~s$^{-1}$ to $\text{TS}_{\text{halo}}=363$ at $V_{PSR}=950~$km~s$^{-1}$. The $\Delta \text{AIC}$ between the halo hypothesis and the overlaid model ranges from $\Delta \text{AIC}=23$ for the $V_{PSR}=300~$km~s$^{-1}$ case, up to $\Delta \text{AIC}=35$ in the $V_{PSR}=0~$km~s$^{-1}$ case, showing that the halo model is significantly preferred over the overlaid model of Gaussian components.

The second row shows similar plots but for realizations of a simulated halo with a higher diffusion coefficient normalization of $D_0=2\times10^{27}~\text{cm}^2~\text{s}^{-1}$, where the emission is significantly more extended. Globally, $\text{TS}_{\text{halo}}$ is lower when the diffusion coefficient is higher. This also leads to a slightly lower $\Delta \text{AIC}$ in favor of the halo hypothesis. 

The third row shows realizations for the case when the pulsar is younger with $P_0=70~$ms (i.e. with a true age of 25~kyr) and $D_0=2\times10^{27}~\text{cm}^2~\text{s}^{-1}$. In comparison with the older case, $\text{TS}_{\text{halo}}$ is lower due to the reduced contribution by relic $e^{\pm}$, ranging from $\text{TS}_{\text{halo}}=90$ at $\text{TS}_{\text{halo}}=136$. The lack of a significant contribution by a relic population of $e^{\pm}$ meant that the emission can be reproduced with a single, less extended Gaussian component and with a harder spectral index than in the older cases, and results in substantially lower $\Delta \text{AIC}$ values.

Fig.~\ref{fig:hess-spectra_high_d} shows an example of simulated and fitted SEDs for the older and younger system with $D_0=2\times10^{27}~\text{cm}^2~\text{s}^{-1}$ and $V_{PSR}=300~$km~s$^{-1}$. The simulated halo and fitted halo spectra are shown, with the latter successfully reproducing the former as expected. The spectra corresponding to the Gaussian components are shown according to their colors in the significance maps (cf. Fig~\ref{fig:hess-maps-all}). When the pulsar is older, two components were required, with the larger component accounting for the contribution of $\gamma$ rays with energies below $E_{\chi}$ by relic $e^{\pm}$, while the harder compact component at the pulsar's position accounts for recently injected $e^{\pm}$. Thus, the detection of a second Gaussian component with a harder spectrum at higher energies implies the detection of an energy-dependent morphology. The sum of the spectra of both Gaussian components slightly underestimates the simulated halo's spectrum below $\sim0.5~$TeV, which might contribute to the preference of the halo hypothesis by the data. For the younger system, only a single component was required to account for the simulated spectrum. The simulated halo spectrum is slightly underestimated below $\sim0.6~$TeV.

\subsection{CTAO sample realizations}
\begin{figure*}[t!]
 \centering
    \includegraphics[width=15.5cm]{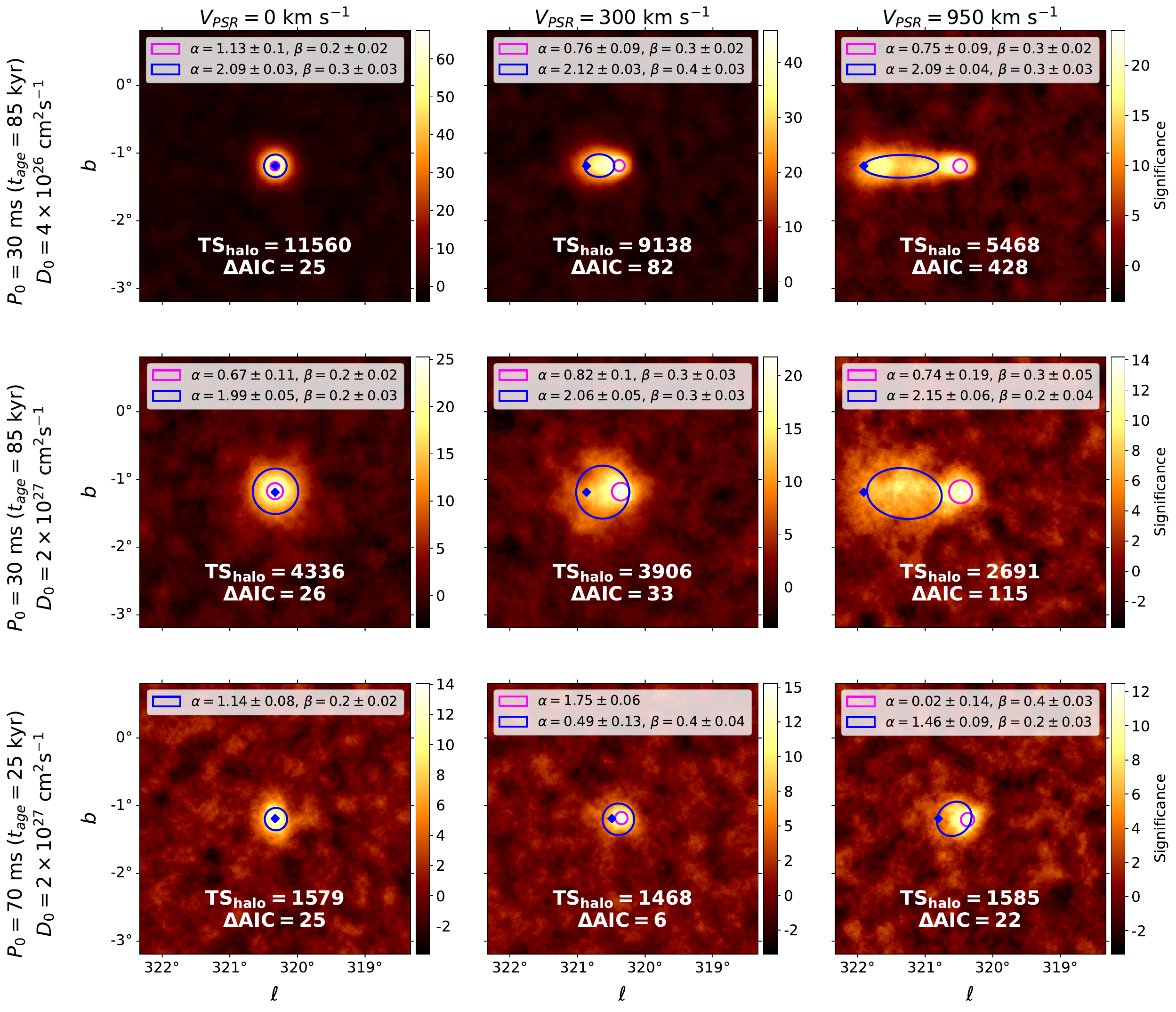} 
  \caption{Same as Fig.~\ref{fig:hess-maps-all}, but with simulated datasets obtained with 42 hours of observations  using 14 MSTs in CTAO-South.}
  \label{fig:ctao-maps-all}
\end{figure*}

As a comparison, Fig.~\ref{fig:ctao-maps-all} shows the significance maps obtained with example realizations using the IRFs of CTAO-South's 14 MSTs. Globally, a dramatic increase in source significance can be seen, with $\text{TS}_{\text{halo}}$ values roughly an order of magnitude higher than those obtained using the H.E.S.S. IRFs. In the older pulsar case, the source is resolved into two components even when $V_{PSR}=0~$km~s$^{-1}$, as opposed to the H.E.S.S. realizations. Additionally, the source is resolved into two components in the younger pulsar case when $V_{PSR}>0~$km~s$^{-1}$. Similar trends to the H.E.S.S. cases are observed, with the source significance increasing with decreasing $D_0$ and increasing pulsar age, and with the more compact components exhibiting significantly harder spectral indices than those of the larger components.

\begin{figure*}[t!]
 \centering
    \includegraphics[width=15.5cm]{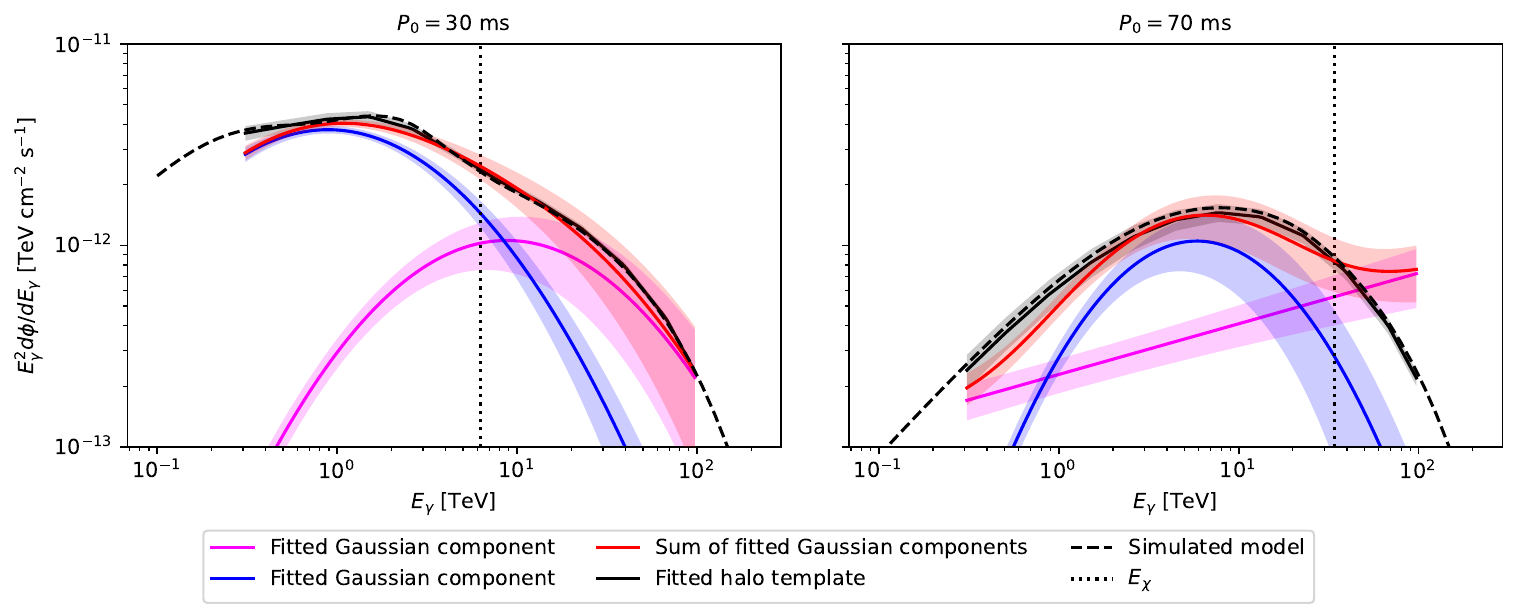} 
  \caption{Same as Fig.~\ref{fig:hess-spectra_high_d} but with datasets simulated using the CTAO-South MST IRFs, for the components of the $D_0=2\times10^{27}~$cm$^2$~s$^{-1}$ and $V_{PSR}=300~$km~s$^{-1}$ cases of Fig.~\ref{fig:ctao-maps-all}.}
  \label{fig:ctao-spectra_high_d}
\end{figure*}

In the first row with the older pulsar $P_0=30~$ms and lower diffusion coefficient $D_0=4\times10^{26}~\text{cm}^2~\text{s}^{-1}$, the $\Delta \text{AIC}$ between the optimal Gaussian model and the halo model ranges from $25$ at $V_{PSR}=0~$km~s$^{-1}$ to $428$ at $V_{PSR}=950~$km~s$^{-1}$. For the latter, the larger component is highly asymmetric, and was not adequately reproduced by the elliptical Gaussian model. Thus, the $\Delta \text{AIC}$ in favor of the halo model is significantly improved in comparison with H.E.S.S., due to the improved sensitivity and angular resolution of CTAO. The second row shows cases with a higher $D_0=2\times10^{27}~\text{cm}^2~\text{s}^{-1}$, where the halo model remains significantly favored with a $\Delta \text{AIC}$ ranging from 26 to 115. The significance of the source is much higher than $5\sigma$, even though it has a similar injection power $\eta L_{\star}$ corresponding to the detection threshold determined by \cite{ecknerCTAhalos} for a halo at $d_{PSR}=3~$kpc. This is because the source of \cite{ecknerCTAhalos} is much less significant primarily due to a higher $B_{ISM}\sim6~\mu\text{G}$ and a softer $\Gamma=2.4$, with softer indices expected to yield less significant detections (at constant injection efficiency) following Fig.~\ref{fig:model-spectra}. The older pulsar halo explored in this work with $D_0=5\times10^{27}~\text{cm}^2~\text{s}^{-1}$ (our closest value to the Geminga-like halo of \citealt{ecknerCTAhalos}), with an injection index $\Gamma=2.4$ and a magnetic field $B_{ISM}=6~\mu\text{G}$, yields $\text{TS}_{\text{halo}}\sim55$. The same halo with $B=3~\mu$G yields $\text{TS}_{\text{halo}}\sim2130$ if $\Gamma=1$, and $\text{TS}_{\text{halo}}\sim830$ if $\Gamma=2$. For reference, we also simulated an identical halo to \cite{ecknerCTAhalos} and found a detection significance $\sim4.2\sigma$, which is comparable to their result ($5\sigma$).

Similarly to Fig.~\ref{fig:hess-spectra_high_d}, Fig.~\ref{fig:ctao-spectra_high_d} shows the fitted spectral models obtained using the CTAO IRFs, corresponding to the older and younger pulsar cases with $V_{PSR}=300~$km~s$^{-1}$ and $D_0=2\times10^{27}~\text{cm}^2~\text{s}^{-1}$. The detection of a spectral curvature for both components at $P_0=30~$ms resulted in a much improved reproduction of the simulated spectrum in comparison with the H.E.S.S. example. Moreover, an energy-dependent morphology was resolved for the younger pulsar case, with two components required to adequately reproduce the simulated spectrum.

\section{Parameter space exploration}\label{sect:param_scan}
\subsection{Characterization of pulsar halos with H.E.S.S.}
\subsubsection{Source significance}
\begin{figure*}[t!]
 \centering
    \includegraphics[width=17cm]{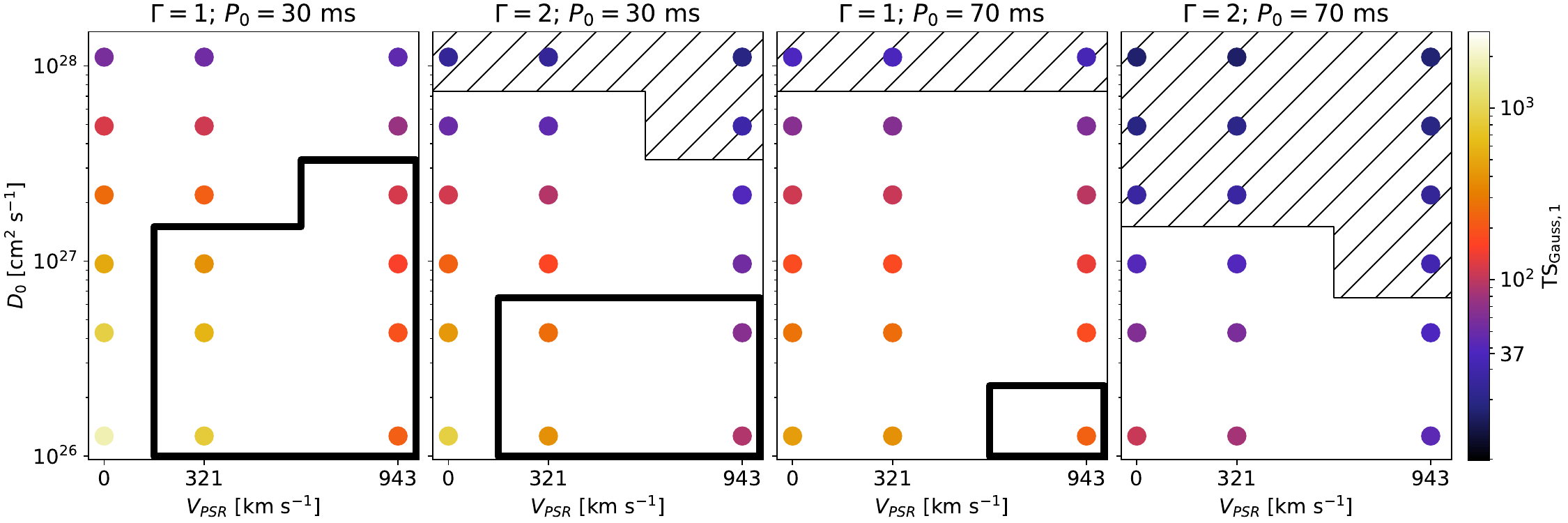} 
  \caption{Average test statistic $\text{TS}_{\text{Gauss}, 1}$ across all realizations of a single fitted 2D Gaussian~$\times$~PL component for different parameter combinations using the H.E.S.S. IRFs with $\sim42$ hours of observation. The hatched region corresponds to $\text{TS}<37$, i.e. the source is not detected. The black contours indicate parameter combinations where a second Gaussian was detected in more than $50\%$ of realizations.}
  \label{fig:hess-ts-gauss1}
\end{figure*}
In this section, the analysis results for the entire pulsar halo simulated parameter space are presented, beginning with the H.E.S.S. simulations. All realizations correspond to $\sim42$ hours of observation time. Traditionally, unidentified extended sources are modeled using a 2D Gaussian$~\times~$PL component. Fig.~\ref{fig:hess-ts-gauss1} shows the average test statistic across all realizations $\text{TS}_{\text{Gauss}, 1}$ where the fit successfully converged, obtained by fitting the data using a single 2D Gaussian~$\times$~PL component. Each panel corresponds to a different combination of the simulated halo's injection spectrum index $\Gamma$ and the pulsar's initial period $P_0$, and shows $\text{TS}_{\text{Gauss}, 1}$ for all simulated values of the diffusion coefficient normalization $D_0$ and pulsar proper motion $V_{PSR}$. The hatched regions cover parameter combinations for which $\text{TS}_{\text{Gauss}, 1}<37$, i.e. lower than the detection threshold of $5\sigma$ for 5 d.o.f. The bold contours encapsulate parameter combinations for which the addition of a second, possibly elliptical, 2D Gaussian component significantly improved the likelihood of the model in more than $50\%$ of realizations. The significant detection of a second Gaussian is generally favored for older systems with a non-zero velocity and a lower $D_0$, because the Gaussian's asymmetry and offset from the pulsar increase. The source's significance is higher when the diffusion coefficient and/or the proper motion are lower, because of the increased surface brightness. The source is undetected for the highest value of $D_0$ across the entire parameter space. Additionally, the detection significance is improved when the injection index is harder (at constant $\eta$), because the SED peaks in the optimal H.E.S.S. energy band. When the pulsar is younger, the addition of a second 2D Gaussian component is not significant, and the source is undetected for the majority of the parameter space with a softer injection index $\Gamma=2$. The 2D Gaussian's $1\sigma$ extension increases predominantly with $D_0$ and $t_{age}$. The older pulsar's extension ranges from $\sim0.08^{\circ}$ to $\sim0.45^{\circ}$ with increasing $D_0$. When $\Gamma=2$ and $V_{PSR}=950~\text{km}~\text{s}^{-1}$, the extension can be roughly twice as high. For the younger pulsar, the extension ranges from $\sim0.05^{\circ}$ up to $\sim0.2^{\circ}$ with increasing $D_0$. The pulsar's proper motion velocity introduces an offset between the pulsar and the Gaussian's centroid, but the pulsar is contained within the Gaussian's $1\sigma$ extent in all cases except for $\Gamma=2$ and $V_{PSR}=950~\text{km}~\text{s}^{-1}$, where the offset ranges from $0.6^{\circ}$ to $1.4^{\circ}$ with decreasing $D_0$.
\subsubsection{Pulsar halo identification}
\begin{figure*}[t!]
 \centering
    \includegraphics[width=17cm]{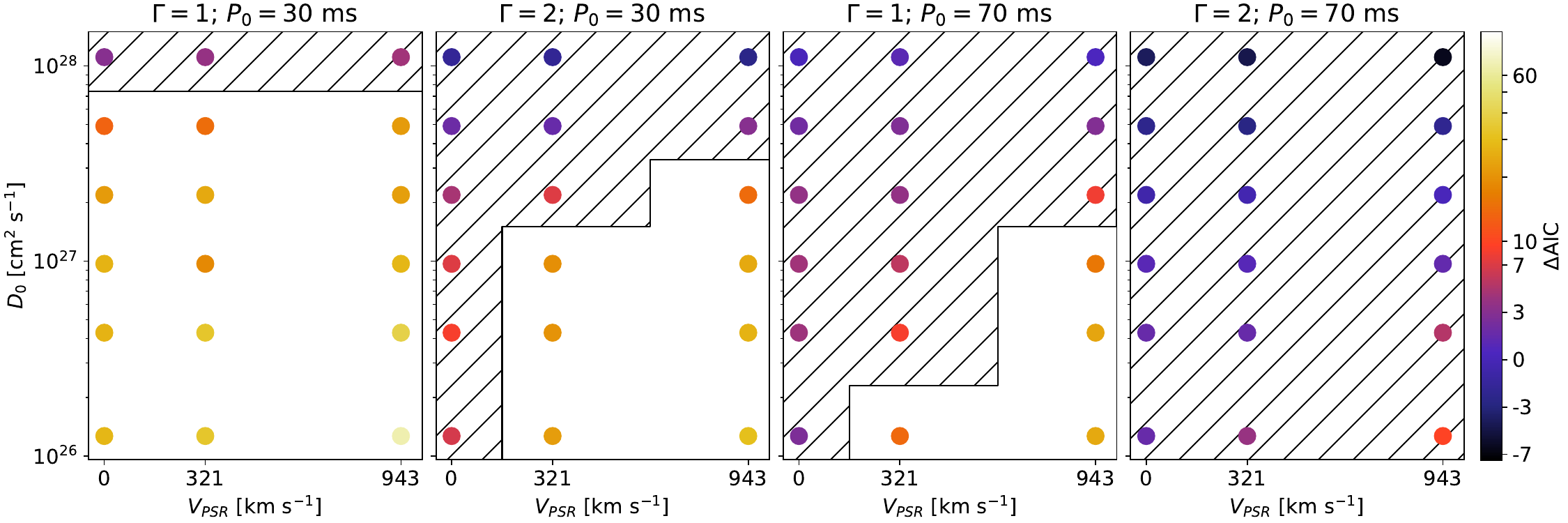} 
  \caption{Average $\Delta \text{AIC}$ across all realizations between the best-fit Gaussian model and halo model for different parameter combinations using the H.E.S.S. IRFs with $\sim42$ hours of observation. The Gaussian model may include two components and curved spectra if significantly detected. The hatched region corresponds to $\Delta \text{AIC}<10$.}
  \label{fig:hess-daic}
\end{figure*}
The ability of the H.E.S.S. array to discriminate between a model of Gaussian components and the physically motivated pulsar halo model was assessed through the Akaike Information Criterion, comparing the $\text{AIC}$ score associated to the pulsar halo template fit with that of the best-fit Gaussian model (i.e. single Gaussian, two Gaussians, or a Gaussian and an elliptical Gaussian). The components of the Gaussian model may also be modeled spectrally with a LP, if it was found to be significant when compared to a PL. The average $\Delta \text{AIC}$ across all realizations for each parameter combination is shown in Fig.~\ref{fig:hess-daic}. The negative $\Delta \text{AIC}$ values mean that on average, the Gaussian model has a lower (better) $\text{AIC}$ score. The hatched region represents $\Delta \text{AIC}<10$, an arbitrary threshold for considering the halo model as being significantly preferred over the Gaussian model. With decreasing $D_0$, $\Delta \text{AIC}$ generally increases due to the rising source significance and due to a stronger peak in the spatial profile at the pulsar's position. It also increases with $V_{PSR}$, because of the prominent asymmetric tail-like feature which can not be adequately reproduced by neither a symmetric nor an elliptical Gaussian profile, despite the source being less significant. Older systems are easier to distinguish from Gaussian models, owing to the presence of a substantial relic population of $e^{\pm}$, resulting in a complex morphology and spectrum that are difficult to reproduce with Gaussian components. When the pulsar is younger, the halo is indistinguishable from a Gaussian model except when the source is sufficiently asymmetric and $\Gamma=1$. The halo model performs poorer when the injection index is softer due to the reduced contribution by higher energy $e^{\pm}$, resulting in a less complex morphology and spectrum that can be accounted for by a single Gaussian component.
\begin{figure*}[t!]
 \centering
    \includegraphics[width=17cm]{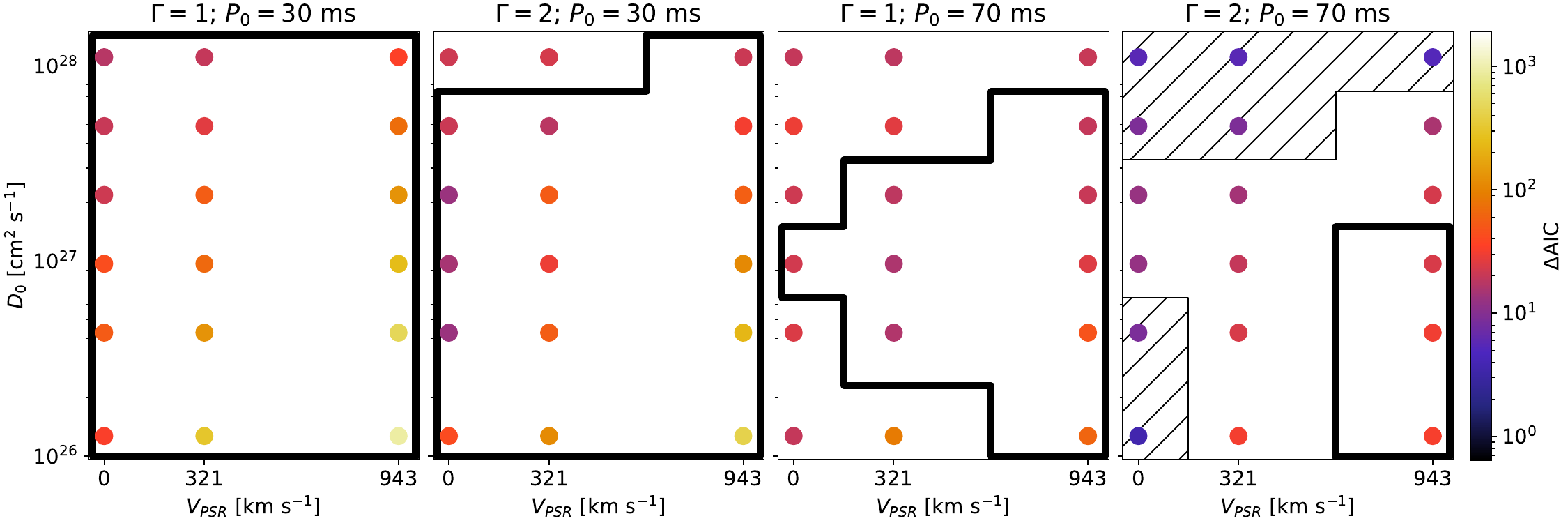} 
  \caption{Same as Fig.~\ref{fig:hess-daic} but with the CTAO-South MST IRFs. The black contours indicate parameter combinations where a second Gaussian was detected in more than $50\%$ of realizations.}
  \label{fig:ctao-daic}
\end{figure*}
\subsection{Pulsar halo identification with CTAO}
With $\sim42$ hours of observation using the CTAO-South MST IRFs, the source was significantly detected (when fitting only one Gaussian component, similarly to Fig.\ref{fig:hess-ts-gauss1} in the H.E.S.S. study) across all the different parameter combinations, and the $\text{TS}_{\text{Gauss},1}$ of the Gaussian sees an improvement of roughly one order of magnitude when compared to the H.E.S.S. simulations, and the same trends were found. Consequently, in this subsection we only focus on the identification of the halo with CTAO.

The average $\Delta \text{AIC}$ values for the CTAO-South MST simulations are shown in Fig.~\ref{fig:ctao-daic}. The black hatches represent $\Delta \text{AIC}<10$, and the bold contours contain parameter combinations for which a second Gaussian was detected at least in $50\%$ of all realizations. In comparison with the H.E.S.S. study, most cases strongly favor the halo model over the Gaussian model, with the exception of a few cases with $\Gamma=2$ and $P_0=70~$ms. Similar trends were generally observed, with the $\Delta \text{AIC}$ increasing in favor of the halo model for a lower $D_0$ and higher $V_{PSR}$, however this is not always followed due to the addition of second Gaussian being favored strongly when $V_{PSR}>0$. Thus, $\sim50$ hours of future observations, with the intermediate array of 14 MSTs at CTAO-South, is ample time for discriminating between the Gaussian model and the halo templates. A second Gaussian is detected in the majority of our realizations, and such a feature in future Galactic surveys could be indicative of the presence of a pulsar halo and/or a component of relic $e^{\pm}$.

\section{Impact of the interstellar emission}\label{sect:GDE}

Modeling the Galactic diffuse emission (GDE) due bremsstrahlung radiation and pion decay from inelastic $p$-$p$ collisions between CRs and interstellar gas can be a source of systematic uncertainties on the measurements of pulsar halo parameters, particularly near the inner Galactic plane. The GDE can be modeled in data analyses with an empirical approach by using maps of interstellar gas tracers. Assuming the cosmic-ray density is constant in the field of view ($4^{\circ}\times4^{\circ}$), the GDE traces the spatial distribution of interstellar molecular and atomic gas. In this work, we used the Planck dust opacity map at 353 GHz \citep{planck_dust}. Specifically, two spatial templates were constructed from two different regions in the map. The first corresponds to a $4^{\circ}\times4^{\circ}$ region centered at Galactic coordinates $(\ell, b)=(30.5^{\circ}, 0^{\circ})$, near W43 (HESS~J1848$-$018). This particular region was chosen due to our past knowledge of an extended structure in the Planck opacity map that allows estimating the impact of the GDE in a region where substantial source confusion with the pulsar halo is expected. The second corresponds to a $4^{\circ}\times4^{\circ}$ region centered at Galactic coordinates $(\ell, b)=(320.3^{\circ}, 0^{\circ})$. The longitude corresponds to the position of MSH~15$-$52, and this portion of the map also contains an extended structure, albeit with a significantly smaller gas density. In both cases, the GDE is included in the simulations by normalizing the dust opacity map (yielding a spatial template in units of steradians$^{-1}$) and multiplying it by a PL with index $\alpha_{\text{GDE}}=2.5$. The PL index value was chosen to be harder than 2.7, similar to the measurements of \cite{gaggero_2015} and \cite{yang_2016} for a pulsar at $\ell=320^{\circ}$ and $6~\text{kpc}$ from the Galactic center (corresponding to $d_{PSR}=3~\text{kpc}$). We tested three PL amplitudes $\phi_{0,\text{GDE}}=2.34\times10^{-11}$, $7.8\times10^{-11}$ and $2.34\times10^{-10}~\text{TeV}^{-1}~\text{cm}^{-2}~\text{s}^{-1}$. These are 3, 10 and 30 times higher than the amplitude for which the $\ell=320.3^{\circ}$ GDE template is detected at the $\sim4.5~\sigma$ level using the H.E.S.S. IRFs, with 42 hours of observation.

\begin{figure*}
 \centering
    \includegraphics[width=14cm]{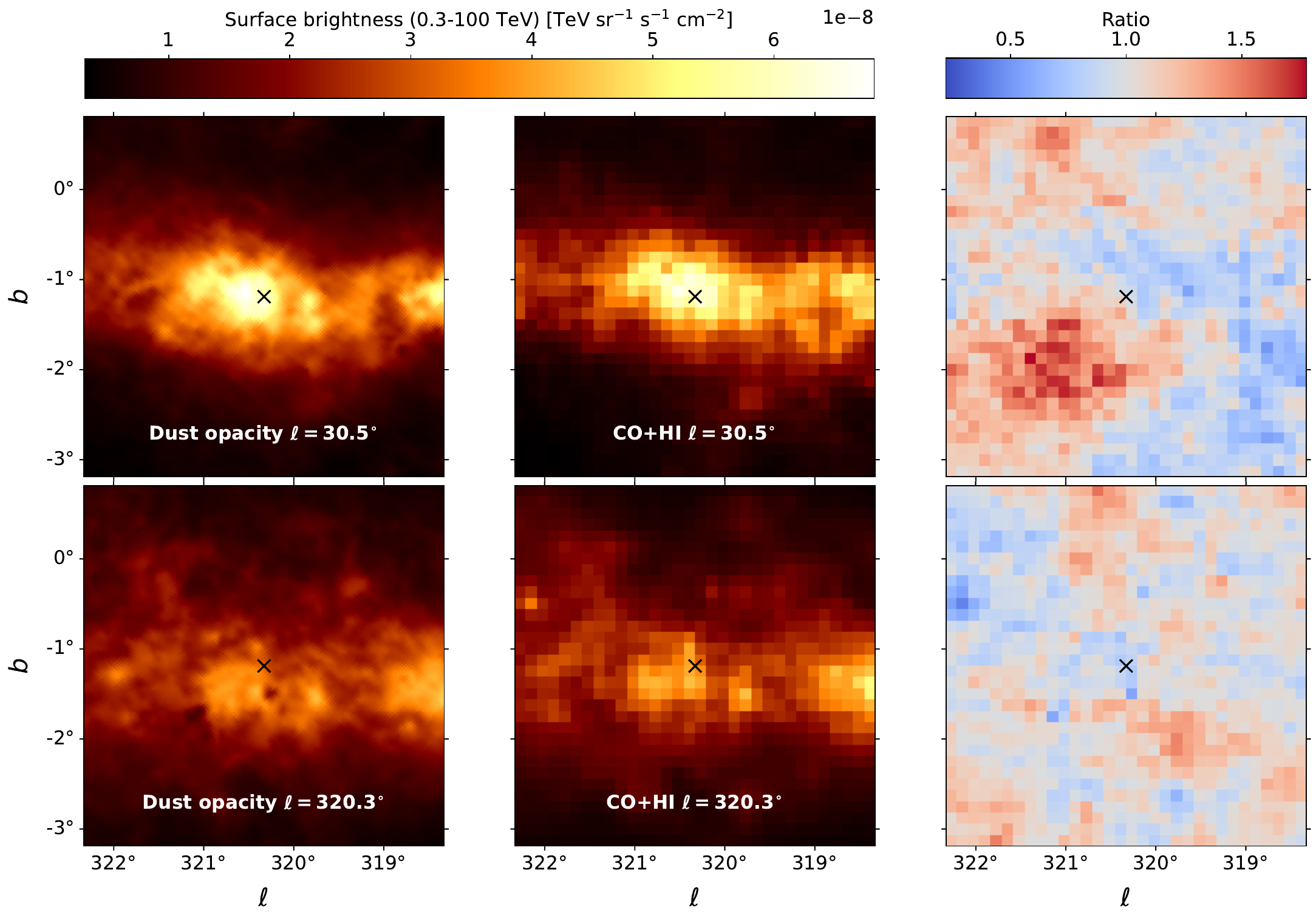} 
  \caption{Surface brightness maps from 0.3 to 100~TeV of the GDE models used for the dataset simulations and analyses. The top row corresponds to the models built from the normalized gas tracer maps at $(\ell, b)=(30.5^{\circ}, 0^{\circ})$, and the bottom row to those at $(\ell, b)=(320.3^{\circ}, 0^{\circ})$. The models' spectral component is a PL with amplitude $\phi_{\text{0,GDE}}=2.34\times10^{-11}~\text{TeV}^{-1}~\text{cm}^{-2}~\text{s}^{-1}$ and index $\alpha_{\text{GDE}}=2.5$. The black cross represents the pulsar's position. \textbf{Left:} GDE models built from the Planck dust opacity map at 353~GHz \citep{planck_dust}. \textbf{Middle:} GDE models built from a combination of the CO intensity map \citep{Dame_2001} and the HI column density map \citep{HI4PI}. \textbf{Right:} Ratio between both maps $\text{Dust}/(\text{CO}+\text{HI})$.}
  \label{fig:gde_maps}
\end{figure*}

For each of the GDE maps that were created, the impact of the GDE on the analysis results for a given realization was assessed in two ways. The first involves fitting the data with the simulated template (and free spectral parameters). The second involves fitting the data with a template constructed from a combination of the $J=1-0$ CO intensity map of \cite{Dame_2001} and the HI column density map of \cite{HI4PI}, to quantify the impact of a potential mismodeling. Fig.~\ref{fig:gde_maps} shows the surface brightness maps from 0.3 to 100~TeV of the two dust-based GDE models that were simulated, and the CO+HI maps that were subsequently used in the fit. Excess $\gamma$-ray emission is expected to trace the structures seen in the ratio between the dust and CO+HI-based templates, which will be compensated for by the pulsar halo model, thus worsening the parameter reconstruction.
\begin{figure*}
 \centering
    \includegraphics[width=17cm]{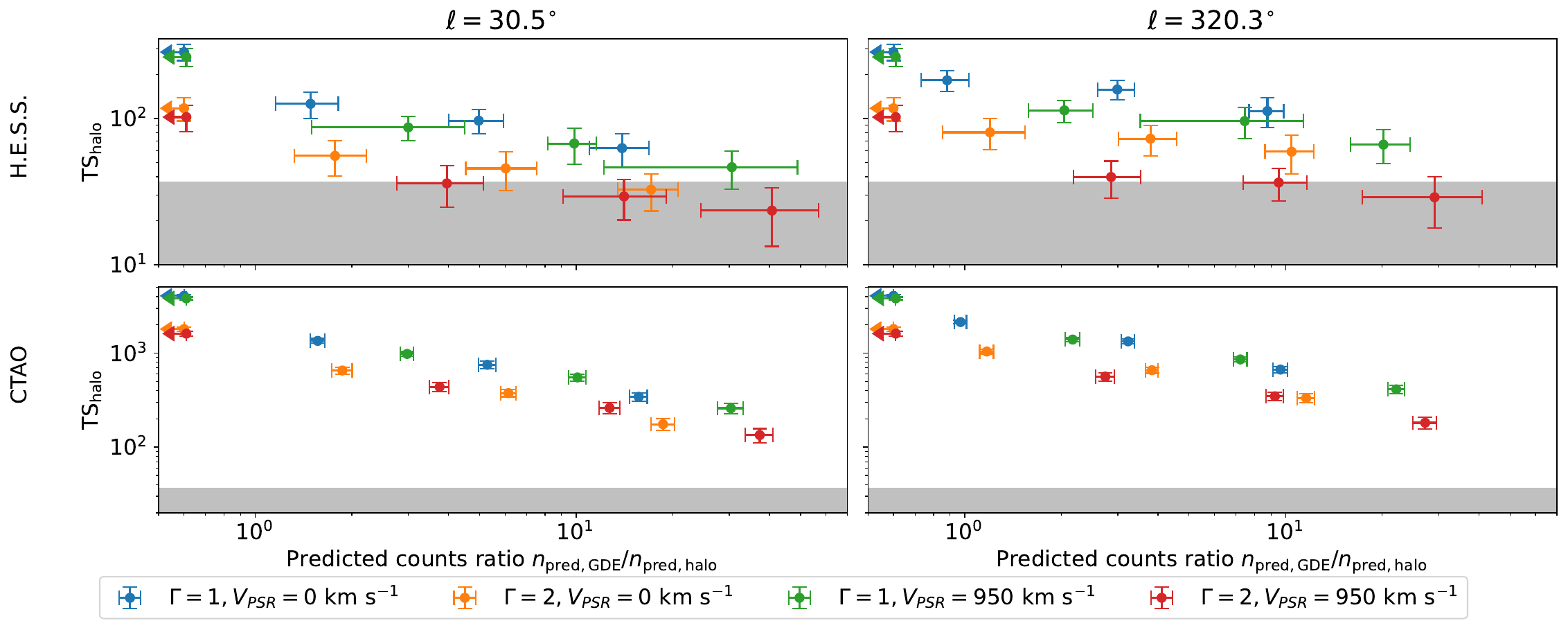} 
  \caption{$\text{TS}_{\text{halo}}$ with respect to the mean ratio between the GDE and halo's predicted counts, across all realizations. The error bars represent the standard deviations. Leftward arrows correspond to $\text{TS}_{\text{halo}}$ when no GDE was simulated nor fitted. Each color represents different simulated pulsar halo parameters, with $D_0=2\times10^{27}~\text{cm}^{2}~\text{s}^{-1}$ and $P_0=30~$ms. The shaded grey area represents the $5\sigma$ detection threshold. The top row and bottom row correspond to the H.E.S.S. and CTAO simulations, while the left and right panels correspond to the GDE templates at $\ell=30.5^{\circ}$ and $\ell=320.3^{\circ}$, respectively.}
  \label{fig:ts_red_npred}
\end{figure*}

The impact of the GDE was assessed for the older pulsar halo with $P_0=30~$ms, $D_0=2\times10^{27}~\text{cm}^{2}~\text{s}^{-1}$, both values of $\Gamma$ (1 and 2) and two values of the proper motion $V_{PSR}=0~\text{km}~\text{s}^{-1}$ and $950~\text{km}~\text{s}^{-1}$. The chosen diffusion coefficient is the highest value for which the $\Gamma=2$ halo is significantly detected when simulated with the $\ell=30.5^{\circ}$ dust map. This study was constrained to these parameters primarily for computational reasons. The exact same analysis procedure was conducted as presented in Section~\ref{sect:simulation}, using both the H.E.S.S. and CTAO IRFs, for 50 realizations in each tested case.

When the GDE is modeled using the same model in the simulation and the fit, the estimated pulsar halo parameters remained consistent with the simulated parameters, albeit with a larger variance across all realizations. The fitted halo model's test statistic $\text{TS}_{\text{halo}}$ is significantly impacted when compared to the analyses without a simulated GDE component, as shown in Fig.~\ref{fig:ts_red_npred}. The average halo's test statistic across all realizations is shown against the average ratio between the GDE model and the halo model's predicted counts, $n_{\text{pred, GDE}}$ and $n_{\text{pred, halo}}$. The model-predicted counts are extracted within a $0.5^{\circ}$ radius around the pulsar's position if $V_{PSR}=0~\text{km}~\text{s}^{-1}$, or in a $2.5^{\circ}\times1^{\circ}$ rectangle extending along the relic halo ``tail'' if $V_{PSR}=950~\text{km}~\text{s}^{-1}$. The reduction in $\text{TS}_{\text{halo}}$ is roughly consistent between H.E.S.S. and CTAO in both the $\ell=30.5^{\circ}$ and $\ell=320.3^{\circ}$ cases, ranging from $\sim30\%$ to $\sim90\%$, depending on the predicted counts ratio and the simulated GDE template.

The impact on the pulsar halo's parameter reconstruction when fitting the simulated GDE with the CO+HI-based templates is shown in Fig.~\ref{fig:gde_param_reco_hess_cohi} with the H.E.S.S. IRFs, for $\Gamma=1$. The results for $\Gamma=2$ and with the CTAO IRFs are highly similar, and therefore not shown. Large discrepancies between the estimated and fitted values manifest with increasing $\phi_{0,\text{GDE}}$, as the excess structures shown in Fig.~\ref{fig:gde_maps} become more significant. The pulsar halo model compensates for the excess, thus leading to biases in the estimated parameters. This can be seen by the estimated direction of the pulsar's proper motion $V_{PSR, \Theta}$. Thus, the impact of mismodeling the GDE is non-negligible for H.E.S.S. and CTAO and detailed systematic studies would be required to properly estimate the parameters in analyses of real data, particularly in regions where substantial contamination by the GDE is expected.
\section{Conclusion}\label{sect:conclusion}
In this work, we assessed the ability of the H.E.S.S. and CTAO-South MST arrays to constrain the parameters of a spatially resolved model of escaped $e^{\pm}$ diffusing in the ISM, injected by a middle-aged PWN, using a template-based modeling approach with the Gammapy library. We tested the model with $\sim42$ hours of simulated observations of pulsar halos, with physical parameters motivated by pulsar halo candidates seen in real data. The morphology and spectra that could potentially be observed with H.E.S.S and CTAO were characterized for the entire parameter space, in addition to the source's detection significance and the potential of the halo template approach to distinguish the source from a model containing 2D Gaussian components. 

For the hypothetical pulsar halos assumed in this work, we found that an older pulsar, lower proper motion velocity and lower diffusion coefficient favor the significant detection of the source. At constant injection efficiency, a harder $e^{\pm}$ injection index is also found to yield a more significant detection.

When modeled with 2D Gaussian components, the source can be resolved into two components with the H.E.S.S. array, if the age and the velocity of the pulsar are high enough and the diffusion coefficient is low enough. In these cases, a diffuse relic component can be resolved, reminiscent of the cases of HESS~J1809$-$193 and HESS~J1813$-$178. The detection of two Gaussian components with CTAO is favored in the majority of the explored parameter space in this work. Finally, we found that the halo can reach extensions $\gtrsim0.5^{\circ}$, similar to some extended components in the HGPS that were discarded due to their large size and low surface brightness (having been deemed as possible artifacts of mismodeling the GDE). These may be revisited, should a robust detection be confirmed in future surveys.

The pulsar halo model is statistically preferred over a model containing 2D Gaussian components in H.E.S.S. data when the pulsar is older and for a harder injection index, owing to the more complex energy-dependent morphology. When the injection index is softer and the pulsar is younger, this is only true for a low enough diffusion coefficient and high enough proper motion. Thus, despite these conditions also favoring the detection of a second Gaussian, the halo template's statistical preference by the data increases. The halo model is statistically preferred with CTAO for nearly the entire explored parameter space. 

Finally, we tested the impact of the GDE by simulating and fitting models constructed from interstellar gas tracer maps. When the GDE is modeled using the same template from which the data was simulated, a significant decrease in the source's significance is found, the value of which depends on the surface brightness ratio between the pulsar halo and the GDE, and the GDE's local morphology at the source. Nevertheless, the pulsar halo's parameters can be adequately reconstructed. When the GDE is modeled using a different spatial template, and if its surface brightness is high enough, a substantial bias was found in the pulsar halo's parameter estimation.

The method can in principle be used with any physically-motivated model, and any datasets that are supported by the Gammapy library, allowing for more precise parameter estimation through multi-instrument joint analyses. A natural prospect is to apply the template-based approach to real data, which could help identify significant-enough pulsar-powered sources. Future works could further our understanding of pulsar halo formation, such as the transition from the PWN stage to the halo stage, and the required environmental conditions.

\begin{acknowledgements}
This work made use of Gammapy \citep{gammapy:2023}, a community-developed Python package. The Gammapy team acknowledges all Gammapy past and current contributors, as well as all contributors of the main Gammapy dependency libraries: \href{https://numpy.org/}{NumPy}, \href{https://scipy.org/}{SciPy}, \href{http://www.astropy.org}{Astropy}, \href{https://astropy-regions.readthedocs.io/}{Astropy Regions}, \href{https://scikit-hep.org/iminuit/}{iminuit}, \href{https://matplotlib.org/}{Matplotlib}. This work made use of Astropy: \footnote{https://www.astropy.org} a community-developed core Python package and an ecosystem of tools and resources for astronomy \citep{astropy:2013, astropy:2018, astropy:2022}. This work made use of data from the H.E.S.S. DL3 public test
data release 1 \citep{hess_public}. This research has made use of the CTA instrument response functions provided by the CTA Consortium and Observatory, see https://www.ctao-observatory.org/science/cta-performance/ (version prod5 v0.1; \citealt{ctao_irfs}) for more details. This work made us of the Sherpa package \citep{sherpa:2001, sherpa:2007, sherpa:2024}.
\end{acknowledgements}

\bibliographystyle{aa}  
\bibliography{refs} 

\begin{appendix}
\onecolumn
\section{Solution to the particle transport equation}\label{app:halo_model}
Following \cite{syrovatskii}, by defining the particle diffusion length scale $\lambda$ and the time since injection $t_{cool}$ as: 
\begin{ceqn}
\begin{equation}
    \lambda^2(E, E')=-4\int_E^{E'} \frac{D(E'')}{\dot{E}(E'')}dE''
\end{equation}
\end{ceqn}
\begin{ceqn}
\begin{equation}
    t_{cool}=t-t'=-\int_E^{E'}\frac{dE''}{\dot{E}(E'')}    
\end{equation}
\end{ceqn}
where $E'(E,t')$ is the particle's energy when injected at a time $t'$ after the pulsar's birth (the pulsar's birth corresponds to $t'=0$). $E'$ is computed for every $t'$ and $E$ by solving $t-t'-t_{cool}=0$. This is only possible if the maximum cooling time
\begin{ceqn}
\begin{equation}
    t_{cool,max}(E)=-\int_E^{+\infty}\frac{dE''}{\dot{E}(E'')}
\end{equation}
\end{ceqn}
satisfies $t_{cool,max}>t-t'$. In other words, the time it takes a particle injected with energy $E'$ at time $t'$ to cool down to an energy $E$ at the time $t$ is limited by $t-t'$. This results in a break in the cooled differential particle density at the energy $E\sim E_{\chi}$ for which $t_{cool,max}(E_{\chi})=t$. Assuming an infinite boundary, a Green's function can be found and the semi-analytical solution to Eq.~\ref{eq:diff_N} reads:
\begin{ceqn}
\begin{equation}
    N(E, r, t)=\int_0^{t} dt' \frac{\dot{E}(E')}{\dot{E}(E)}Q(E', t')\frac{\exp{-\frac{|\mathbf{r}-\mathbf{r}_s(t')|^2}{\lambda^2(E,E')}}}{\pi^{3/2}\lambda^3(E,E')} \label{eq:sol_N}
\end{equation}
\end{ceqn}
Setting the origin of the reference frame $(\mathbf{r}=\mathbf{0})$ at the pulsar's present-day position, the pulsar's transverse velocity in the sky $\mathbf{V}_{PSR}$ may be incorporated by defining $\textbf{r}_s$ in the $-\mathbf{V}_{PSR}$ direction with $r_s(t')=(t-t')V_{PSR}$ (e.g. \citealt{DiMauro2019}). The solution assumes a cylindrical symmetry about the axis in the $\mathbf{V}_{PSR}$ direction. In cylindrical coordinates with the unit vectors $\mathbf{u}_{\rho}$ taken along the cylinder's radius and $\mathbf{u}_{z}$ taken along the transverse velocity axis (i.e. the cylinder's height):
\begin{ceqn}
\begin{eqnarray*}
|\mathbf{r}-\mathbf{r}_s(t')|^2 &=& |\rho\mathbf{u}_{\rho}+z\mathbf{u}_{z}-(t-t')V_{PSR}\mathbf{u}_{z}|^2 \\
    &=& \left(z-(t-t')V_{PSR}\right)^2 + \rho^2 \\
    &=& z^2+(t-t')^2V_{PSR}^2-2z(t-t')V_{PSR}+\rho^2
\end{eqnarray*}
\end{ceqn}
The $\gamma$-ray intensity is computed by projecting the differential particle density distribution $N$ onto the sky, i.e. by integrating $N$ over the line of sight $s$ from 0 to $+\infty$. In the cylindrically symmetric case, $z=\theta d_{PSR}$ and  $\rho^2=\psi^2 d_{PSR}^2+s^2$ (by taking $\mathbf{u}_{z}$ and $\mathbf{u}_{\rho}$ perpendicularly to the line of sight), where $\theta$ and $\psi$ are the angular separations from the pulsar's present-day position. $\theta~\text{and}~\psi\ll1$, therefore the lines of sight for all $\theta$ and $\psi$ are assumed to be parallel. The projected $e^{-/+}$ distribution per unit energy per steradian reads:
\begin{ceqn}
\begin{equation}
    P(E, \theta, \psi,  t)=d_{PSR}^2 \int_0^{t} dt' \frac{\dot{E}(E')}{\dot{E}(E)}Q(E', t')\frac{\exp{\left(-\frac{\theta^2d_{PSR}^2+\psi^2d_{PSR}^2+(t-t')^2V_{PSR}^2-2\theta d_{PSR}(t-t')V_{PSR}}{\lambda^2(E,E')}\right)}}{\pi\lambda^2(E,E')}
\end{equation}
\end{ceqn}
At each position $(\theta, \psi)$, the differential photon spectrum due to ICS from the particle distribution $P$ is computed using the Naima package \citep{naima}. This yields the $\gamma$-ray intensity $I(E_{\gamma}, \theta, \psi)$ with $E_{\gamma}$ the photon energy. If $V_{PSR}=0$, the solution assumes a spherical symmetry and the intensity reduces to $I(E_{\gamma}, \zeta)$ with $\zeta^2=\theta^2+\psi^2$. 
\section{Pulsar halo parameter reconstruction}\label{app:param-rec}
\begin{figure*}[ht!]
\centering
   \includegraphics[width=17cm]{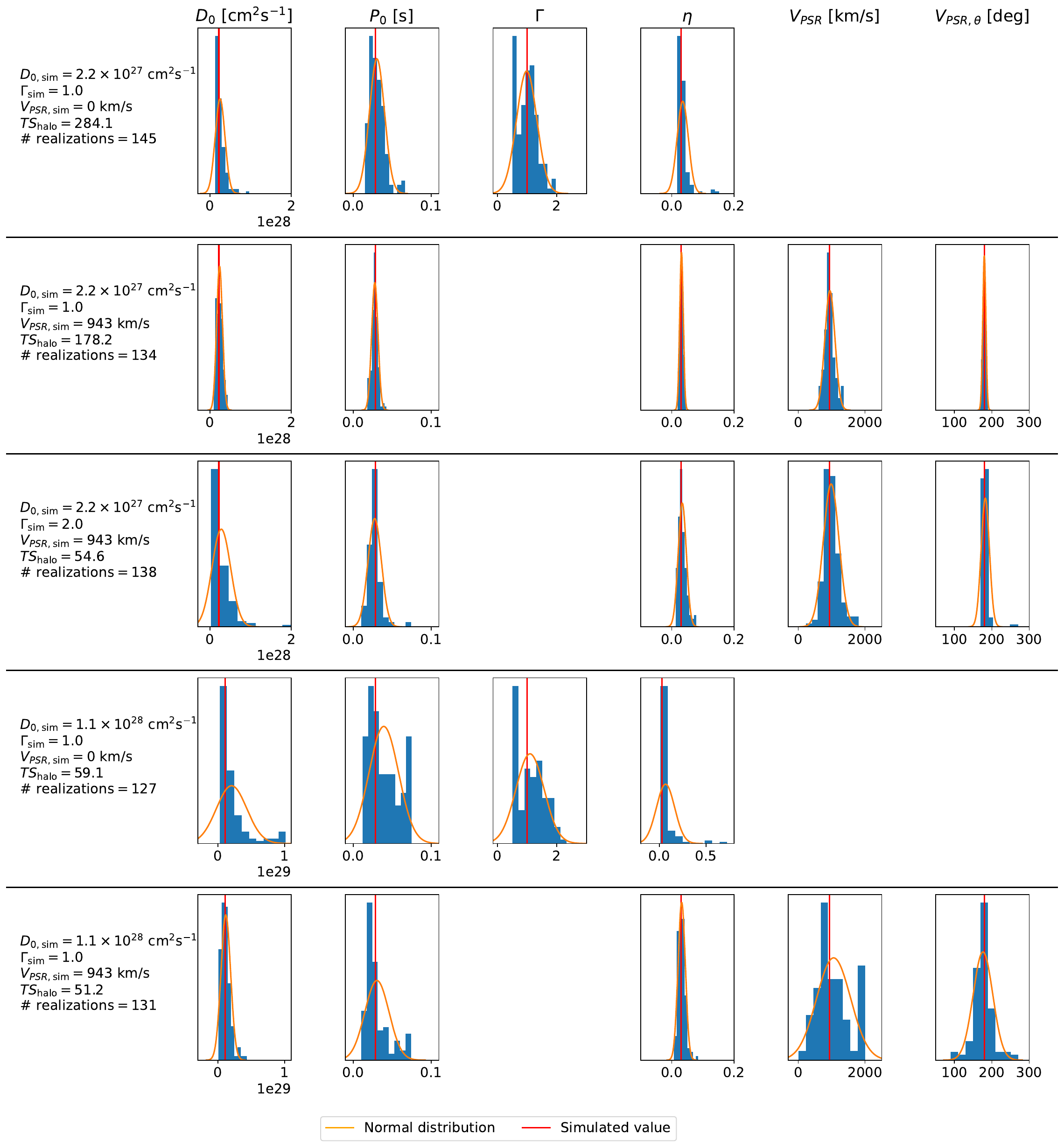}
     \caption{Parameter recovery for six different parameter combinations, each corresponding to one row. The histograms are the density distributions of the estimated values across all realizations. $\text{TS}_{\text{halo}}$ corresponds to the average Test Statistic of the fitted halo model across all realizations for the given parameter combination. The red line indicates the simulated value, and the yellow distribution is a normalized Gaussian with the mean and standard deviation of the estimated values across all realizations. For all six rows, $P_0=30~$ms and $\eta=3\%$. All the simulations are generated from $\sim42$ hours of observation. The number of successful realizations is indicated on each row.}
     \label{fig:hess-param-recovery-example}
\end{figure*}

Fig.~\ref{fig:hess-param-recovery-example} shows an example of parameter reconstruction for five different simulated parameter combinations with $P_0=30~$ms, $\eta=3\%$, and different values of $D_0$, $\Gamma$ and $V_{PSR}$ in each row. The histograms represent the density of the estimated value in each realization. The red line represents the simulated value and the yellow line is a normal distribution with the mean and standard deviation of the estimated value across the realizations. When the source is not too significant, some of the realizations do not converge successfully in the fit, and these realizations are excluded from the statistics (the number of successful realizations is indicated for each parameter combination). The mean Test Statistic of the halo fit across all realizations ($\text{TS}_{\text{halo}}$) is indicated for each parameter combination. The template fitting procedure yields convenient estimations of the simulated parameter values. All simulations in this work are generated from $\sim42$ hours of observation. We found that the parameter reconstruction with the CTAO-South MST arrays is much improved, with significantly narrower distributions.
\pagebreak
\\
\\
\\
\\
\\
\\
\\
\section{Modeling the dust-based Galactic diffuse emission with the $\text{CO}+\text{HI}$ template}\label{app:iem}
\begin{figure*}[h!]
 \centering
    \includegraphics[width=17cm]{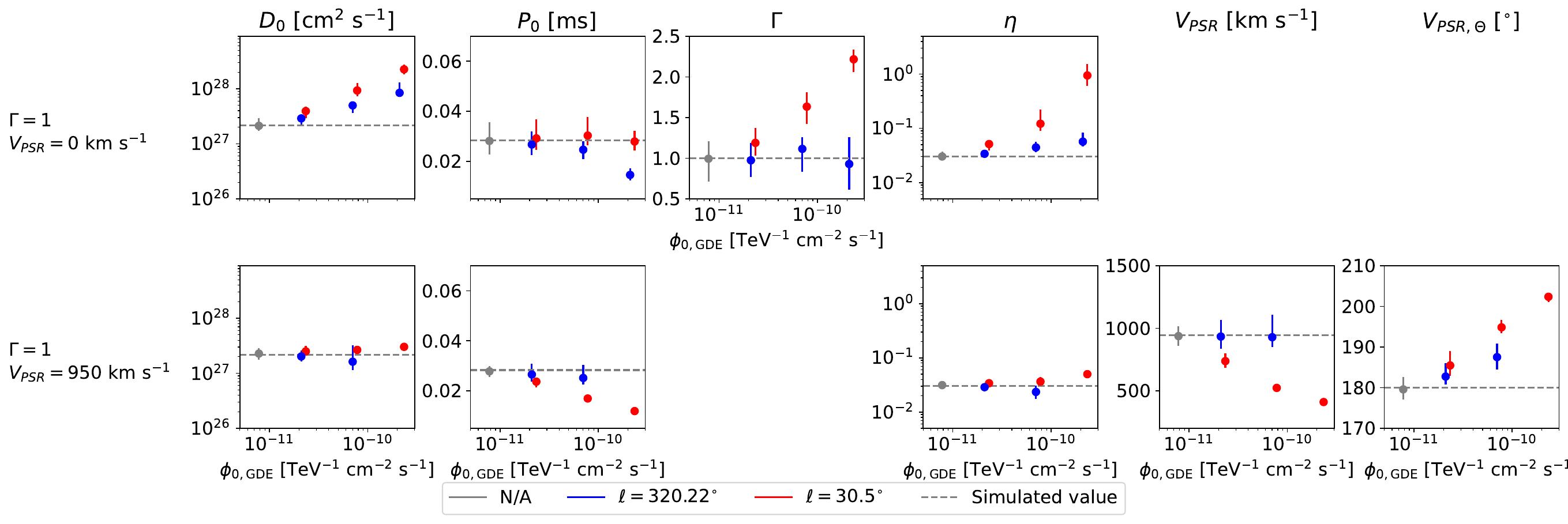} 
  \caption{Pulsar halo parameter reconstruction when simulating a GDE component based on the dust map and fitting a GDE component based on the CO+HI map. Three different values of the simulated GDE's spectral amplitude $\phi_{0, \text{GDE}}$ are shown. Each row corresponds to a different parameter combination, with $\eta=3\%$. Each column corresponds to a fitted halo parameter. The horizontal dashed line shows the simulated value. Each data point represents the median estimated value across all realizations, and the error bars correspond to the first and third quartiles. The grey data represents the case where no GDE was simulated nor modeled. The blue data corresponds to realizations that include a GDE component produced from the dust map at $\ell=320.3^{\circ}$, and the red data those corresponding to $\ell=30.5^{\circ}$.}
  \label{fig:gde_param_reco_hess_cohi}
\end{figure*}

Fig.~\ref{fig:gde_param_reco_hess_cohi} shows the average estimated values of the pulsar halo model's parameters across all realizations, when the GDE component is modeled using the templates constructed from the combined CO and HI maps. Each row represents a different parameter combination, and each column a different pulsar halo parameter. In some cases, the estimated parameter values' distribution was not Gaussian, and we therefore used the median, the first and the third quartiles, the latter of which correspond to the asymmetric error bars. Cases where the optimization converges towards the halo model's parameter limits in a substantial number of realizations are discarded. For reference, the parameter estimates when no GDE was simulated nor fitted is included in each panel.

\end{appendix}
\end{document}